\documentclass[aps,amsmath, amssymb,12pt,letterpaper]{article}
\usepackage{jheppub}

\usepackage{graphicx}
\usepackage{caption} 
\usepackage{subcaption} 
\usepackage{bm}
\usepackage{amssymb}
\usepackage{amsmath}
\usepackage{cancel}
\usepackage{epigraph}

\newcommand{\pref}[1]{(\ref{#1})}

\newcommand{\hk}{hyperK\"{a}hler }
\newcommand{\ds}{de~Sitter }

\newcommand{\ket}[1]{|#1 \rangle}

\def\({\left(}
\def\){\right)}
\def\[{\left[}
\def\]{\right]}

\def\ie{{\it i.e.}}

\def\etc{{\it etc}}

\def\barray{\begin{array}}
\def\earray{\end{array}}
\def\be{\begin{equation}}
\def\ee{\end{equation}}
\def\ben{\begin{equation} \nonumber}
\def\een{\end{equation}}
\def\bea{\begin{eqnarray}}
\def\eea{\end{eqnarray}}
\def\eal{\end{align}}
\def\bal{\begin{align}}

\def\sst{\scriptscriptstyle}
\def\({\left(}
\def\){\right)}
\def\half{\frac12}
\def\coeff#1#2{{\textstyle \frac{#1}{#2}}}
\def\hf{\coeff12}

\def\One{{\hbox{ 1\kern-.8mm l}}}

\def\AA{{\mathcal{A}}}
\def\BB{{\mathcal{B}}}
\def\FF{{\mathcal{F}}}

\def\II{{\mathcal{I}}}
\def\LL{{\mathcal{L}}}
\def\MM{{\mathcal{M}}}

\def\OO{{\mathcal{O}}}

\def\SS{{\mathcal{S}}}

\newcommand{\bC}{{\mathbb C}}

\newcommand{\bR}{{\mathbb R}}
\newcommand{\bS}{{\mathbb S}}
\newcommand{\bT}{{\mathbb T}}
\newcommand{\bZ}{{\mathbb Z}}

\def\ehat{\hat e}

\def\bark{\bar k}

\def\cc{{\bf c}}
\def\xx{{\bf x}}
\def\yy{{\bf y}}
\def\X{{\bf X}}

\newcommand{\sr}{\mathsf r}
\newcommand{\sy}{\mathsf y}
\newcommand{\sH}{\mathsf H}
\def\LdB{{\mathsf L}}
\newcommand{\sQ}{\mathsf Q}

\def\ext{{\rm ext}}

\def\lpl{\ell_{\rm pl}}

\def\lstr{\ell_{\rm s}}
\def\gstr{g_{\rm s}}
\def\gym{g_{\rm \scriptscriptstyle YM}}

\begin{document}


\title{%
The Cheshire Cap
%
}
\author{Emil J. Martinec}

\affiliation{Enrico Fermi Institute and Department of Physics, University of Chicago\\ 5640 S. Ellis Ave., Chicago, IL 60637-1433, USA}

\emailAdd{e-martinec@uchicago.edu} 

\abstract{
A key role in black hole dynamics is played by the inner horizon; most of the entropy of a slightly nonextremal charged or rotating black hole is carried there, and the covariant entropy bound suggests that the rest `floats' in the region between the inner and outer horizon.
An attempt to match this onto results of the microstate geometries program suggests that a `Higgs branch' of underlying long string states of the configuration space realizes the degrees of freedom on the inner horizon, while the `Coulomb branch' describes the black hole exterior; the inter-horizon region has excitations from both branches.  Support for this proposal comes from an analysis of the way singularities develop in microstate geometries, and their close analogy to corresponding structures in fivebrane dynamics.  These singularities signal the opening up of the long string degrees of freedom of the theory, which are partly visible from the geometry side.  A conjectural picture of the black hole interior is proposed, wherein the long string degrees of freedom resolve the geometrical singularity on the inner horizon, yet are sufficiently nonlocal to communicate information to the outer horizon and beyond.  
}

\maketitle


\section{Introduction}

\setlength{\epigraphwidth}{.6\textwidth}

\epigraph{\textit{``Well! I've often seen a cap without a black hole," thought Alice; ``but a black hole without a cap! It's the most curious thing I ever saw in all my life!"}}{\hskip 2.5cm L. Carroll, \textit{Through the Horizon,\\ and what Alice Found There}}

There are by now a variety of constructions of black hole states from a dual field theoretic perspective (for a review, see for example%
~\cite{Aharony:1999ti}).  
In these constructions, nonlocality abounds, perhaps to the degree that typically the field theory is not the place to seek a resolution of puzzles involving local bulk geometry.  Maximally supersymmetric Yang-Mills theory in dimensions $d\le 4$ constructs black holes as thermal states of a brane gas; noncommutativity is a hallmark of the black hole phase of the field theory, and field theory quantum fluctuations are large; at the same time this renders the bulk geometrical description obscure -- how does one reconstruct the bulk geometry?  Where is the black hole horizon, let alone its interior?  Or for that matter any localized bulk observables?  Is the gauge theory only describing the black hole exterior?  Is there some complementarity map to describe the interior%
~\cite{Susskind:1993if}?  What about the experience of infall?   There have been some attempts to construct local observables using the operator spectrum of the gauge theory, dating back to the early days of gauge/gravity duality%
~\cite{Banks:1998dd,%
Balasubramanian:1998sn%
},
but their status is unclear (see%
~\cite{Harlow:2014yka} 
for a recent summary and discussion of the issues).
The recent firewall debate instigated by~\cite{Almheiri:2012rt} (see~\cite{Braunstein:2009v1,Mathur:2009hf} for earlier work) has simply sharpened these issues.  

One would like a testbed which can exhibit as much of the quantum structure of black holes as possible within a setting where the geometry is both weakly coupled and accessible.
Maximally supersymmetric gauge theory is perhaps not that setting, nor is the symmetric orbifold CFT dual to $AdS_3\times \bS^3\times \MM_4$ (where $\MM_4 = \bT^4$ or $K3$).  These field theory realizations of quantum gravity are too unwieldy to answer the sorts of questions one wishes to ask about quasi-local bulk physics; being a weak/strong coupling duality, the weakly coupled regime of the field theory is where the geometry strongly fluctuates, and vice versa.  

At the same time, there has been remarkable progress in elaborating the structure of individual horizonless solutions of effective supergravity equations of motion, beginning with%
~\cite{Lunin:2001fv,Lunin:2002bj} (see%
~\cite{Bena:2013dka} 
for a recent discussion and further references), many carrying the same quantum numbers as BPS black holes, having a large degeneracy, and having the same gap to small excitations.  The suggestion is that these solutions are microstates in the ensemble of states contributing to the black hole entropy.  Proponents hope that the enumeration of these BPS configurations may be nearing an ability to count a substantial fraction of the entropy as a function of the charges and thus, one might hope that these states comprise a dense sampling of the set of microstates.  In addition, there is a sparse but growing set of examples of non-BPS configurations%
~\cite{Jejjala:2005yu,%
Bena:2009qv,%
Compere:2009iy,%
Dall'Agata:2010dy,%
Bobev:2011kk,%
Niehoff:2013mla%
}.

It has not yet been clear how generic these microstates are, and whether one will be able to use them to answer fundamental questions about the flow of information in the course of black hole formation and evaporation.  It seems that there must be some violation of local quantum field theory on macroscopic scales in order for unitarity to be preserved, since the solutions are based on supergravity, for which the classical effective theory obeys the strong energy condition, while Mathur has argued~\cite{Mathur:2009hf} that small corrections to the dynamics cannot solve the information paradox.  One would like to identify the mechanism responsible for this violation of locality and causality on macroscopic scales.  

Techniques have been developed for enumerating large numbers of microstate geometries for particular BPS states carrying three charges, dipole charges and angular momentum, in asymptotically $AdS_3\times \bS^3\times \MM_4$ spacetimes, where $\MM_4$ is $\bT^4$ or $K3$.  We review these constructions in section~\ref{sec:MicroGeom}. The reason to focus on this particular situation is that there are BPS black holes whose horizon is smooth and macroscopic, and so one might hope that supersymmetry nonrenormalization might exert some influence on controlling quantum effects, especially in the near-BPS regime.  These microstate geometries have a number of features that seem generic, including the appropriate gap to non-extremal excitations; and including also the geometry apart from a small region near the horizon, where they differ and roll over to a smooth cap without horizon.  There is a mechanism to support this capping off of an extensive set of geometries somewhat outside the would-be horizon, and to prevent this structure from falling into the black hole, via the interaction of charged sources, fluxes, topology, and geometry.  One is approaching an enumeration which it is hoped can account for a finite fraction of the black hole entropy%
~\cite{Bena:2006is,Warner:2008ma,Bena:2013dka,Bena:2014qxa}, 
however the key question remains of whether a significant fraction of the BPS microstate degrees of freedom can be realized geometrically, and whether the solutions found so far are sufficiently generic.  If these issues coud be resolved favorably, one may imagine that these geometries provide a picture of horizon structure, at least for BPS black holes, and perhaps point us toward resolutions of some of the perplexing puzzles that persist.  The locality/causality issue looms large, however.

In essence, the microstates geometry program asks whether one needs the whole apparatus of nonperturbative gravity in order to understand the quantum structure of black holes; in particular, whether the nonlocalities that are fundamental to a resolution of the information problem and the infall problem, can be discerned from the supergravity approximation.  If so, then one should be able to make considerable progress without a complete nonperturbative description of all of spacetime, instead focussing on the features of the near-horizon structure relevant to the black hole.  It seems as though in trying to reconstruct black hole physics from a complete nonperturbative dual one is working too hard -- first one has to reconstruct all the vastness of AdS, then one has to put a modest size black hole in the middle of it, and distinguish not only one from the other, but specific local features of geometry near the black hole.  The hope is that one can separate this hugely complicated spacetime reconstruction problem from the specific features needed to resolve the puzzles of black holes.

We begin to address these issues in section~\ref{sec:Thermo} in the context of three charge string theory backgrounds and the BTZ black hole geometries they are related to.  The thermodynamic properties of the BTZ geometry together with the covariant entropy bound%
~\cite{Bousso:1999cb,%
Bousso:1999xy,%
Flanagan:1999jp,%
Bousso:2014sda}
point to where the degrees of freedom of the black hole are located; for modest excursions from extremality, most of them reside at the inner horizon.  Section~\ref{sec:MicroGeom} then summarizes the state of the art for constructing the relevant microstate geometries, following%
~\cite{Warner:2008ma,%
Bena:2008dw,%
Niehoff:2013kia,%
Gibbons:2013tqa%
}, 
and section~\ref{sec:Quivers} gives an overview of a complementary approach via quiver quantum mechanics%
~\cite{%
Denef:2002ru,%
Denef:2007vg,%
deBoer:2008fk,%
deBoer:2008zn,%
Manschot:2010qz,%
Lee:2011ph,%
Bena:2012hf%
}.  
The geometry approach matches up quite well with the Coulomb branch of the quiver construction, however the Higgs branch of the quiver has substantially more entropy, leading one to ask where the Higgs branch might be on the geometry side.  Consideration of this question in section~\ref{sec:Fivebranes} leads us to an answer satisfyingly similar to well-understood features of fivebrane dynamics%
~\cite{Kutasov:1995te,%
Aharony:1998ub,%
Giveon:1999zm,
Giveon:1999px,%
Giveon:1999tq,%
Harvey:2001wm%
},
wherein the Coulomb branch has a smooth geometrical cap to the fivebrane throat which can be probed by supergravity at low energies, but which gives way to the Higgs branch of little string theory as the throat grows deeper and the supergravity description becomes singular -- and it is the little string on the Higgs branch which carries the entropy of nonextremal fivebranes.  We identify the analogous structures in the microstate geometries, and argue that the physics is much the same -- that the microstate geometries with deep throats are descriptions of the Coulomb branch near but below the BTZ black hole threshold, and that the Higgs branch opens up the sector of `long strings' that carry the entropy of BTZ black holes~\cite{Strominger:1996sh,Horowitz:1996fn}.
The primary difference is that the tension of little strings is fractionated by a factor of the fivebrane charge quantum $n_5$, while the long strings of $AdS$ are fractionated by a factor $n_1n_5$.
We interpret this result in section~\ref{sec:Discussion} in the light of similar features of matrix theory%
~\cite{Horowitz:1997fr,%
Li:1998ci%
}, 
and propose a mechanism for the resolution of the information paradox~\cite{Hawking:1976ra,Mathur:2009hf} and the associated firewall problem~\cite{Almheiri:2012rt,Braunstein:2009v1} using the interplay of short and long strings, and Higgs and Coulomb branch properties.  A key feature of this scenario is the observation that the long string is at the correspondence point for strings propagating in $AdS_3$~\cite{Giveon:2005mi}, where black holes leave the spectrum; this feature leads one to suspect that the long string simply doesn't see the same geometry that short strings do; that the environment it does experience has no horizon or singularity, and that this property is the basic mechanism by which string theory and its fractionated brane structures resolve null and spacelike singularities in general relativity. 

Thus we find that, like Lewis Carroll's
mischievous creature,%
\footnote{With apologies for the apocryphal epigraph above.}
the cap can disappear at will, but as it fades away into the horizon, it leaves something behind to surprise us. 


\section{Branes, horizons and thermodynamics}
\label{sec:Thermo}

\subsection{Branes and black holes}

Our focus will be the set of three-charge geometries in toroidally compactified string theory.  The full non-extremal D1-D5-P geometry carrying all possible charges, dipole charges and angular momenta is somewhat complicated.  To begin, let us specialize to backgrounds carrying no dipole charges or transverse angular momentum%
~\cite{Horowitz:1996fn}; 
these will illustrate some of the main features we wish to explore:
\bea
ds^2 &=& (H_1 H_5)^{-1/2}\Bigl[ -dt^2+dz^2 + (1-f)(\cosh\alpha_p \,dt +\sinh\alpha_p\, dz)^2 \Bigr]
\nonumber\\
& & + (H_1 H_5)^{1/2}(f^{-1}d\rho^2 + \rho^2 d\Omega_3^2) + \Bigl( \frac{H_1}{H_5} \Bigr)^{1/2} ds^2_{\bT^4}
\eea
where
\be
H_{1,5} = 1+\frac{\rho_0^2 \sinh^2\alpha_{1,5}}{\rho^2}
\quad,\qquad
f = 1-\frac{\rho_0^2}{\rho^2}
\ee

The decoupling limit of D1-D5-P bound states takes 
$\lstr,\rho_0\to 0$ with 
$Q_{1,5,p} = \rho_0^2 \sinh 2\alpha_{1,5,p}$ 
fixed, in such a way that the D1 and D5 charges make a `heavy' background geometry whose contributions to the ADM mass scale like $\lstr^{-2}$, and the P charge is comprised of `light' excitations on that background whose energies scale like $\lstr^0$.
This limit leads to a geometry that is locally $AdS_3\times \bS^3 \times \bT^4$:
\bea
ds^2= \frac{1}{\sqrt{H_1 H_5}}\bigl[-dt^2+dz^2 + H_p(dt + dz)^2\bigr]
+ \sqrt{H_1 H_5}\bigl(f^{-1}d\rho^2 + \rho^2 d\Omega_3^2\bigr) + \sqrt{ \frac{H_1}{H_5} } ds^2_{\bT^4}
\nonumber\\
\eea
where
\be
H_{1,5,p} = \frac{Q_{1,5,p}}{\rho^2}
\quad,\qquad
f = 1-\frac{\rho_0^2}{\rho^2}
\ee
The canonical BTZ form of the metric arrives upon making the coordinate transformation (defining the $AdS$ radius $\ell=4G_3n_1n_5$ in 3d Planck units, where $n_{1,5}$ are the integer brane charges)
\be
r^2 =  \frac{\rho^2}{\ell^2} + \frac{\rho_0^2}{\ell^2} \sinh^2 \alpha_p
\ee
which recasts the $(t,z,r)$ part of the metric locally in the form of a 3d BTZ black hole
(see for example~\cite{Carlip:1995qv} for a review)
\bea
\label{BTZgeom}
ds^2 &=& -N^2 dt^2 + N^{-2}{dr^2} +r^2(dz-N_\varphi dt)^2+\ell^2 d\Omega_3^2
\nonumber\\
N^2 &=& \frac{r^2}{\ell^2} - M_3 + \frac{16 G_3^2 J_3^2}{r^2}
\\
N_\varphi &=& \frac{4 G_3 J_3}{r^2} ~,
\nonumber
\eea
where $J_3$ is the integer momentum charge $n_p$.
Rotation on the three-sphere transverse to the branes fibers the $\bS^3$ over the locally $AdS_3$ BTZ base~\cite{Cvetic:1998xh}.


\subsection{Horizons and thermodynamics}
\label{sec:HorizonThermo}

We will denote the BTZ black hole and similar geometries with stationary horizons as {\textbf{\emph {ensemble geometries}}}, in that the properties of their horizon(s) determine the thermodynamics of the system.
The ensemble geometry should reflect certain average characteristics of the individual microstates, and thus represents a sort of mean field theory for the full dynamics.  The picture is good for motion of macroscopic observables but misses the evolution of quantum correlations in individual microstates.  The question is how to recover the latter while not disturbing the former or leading to gross violations of causality over macroscopic distances, and thus resolve the issue of unitarity of black hole evaporation from the geometrical side of the gauge/gravity duality.

For example, in the BTZ example, the two roots $r_\pm$ of the vanishing $N=0$ of the lapse function~\pref{BTZgeom} are the locations of the inner and outer horizons of the ensemble geometry:  
\be
r_\pm^2 = \frac{M_3\ell^2}{2}\left[ 1\pm \Bigl[1-\Bigl(\frac{8G_3J_3}{M_3\ell}\Bigr)^2\Bigr]^{1/2}\right]
\ee
or equivalently
\be
M_3 = \frac{r_+^2+ r_-^2}{\ell^2}
~~,~~~
J_3 = \frac{r_+ r_-}{4G_3\ell} ~;
\ee
the thermodynamic variables of the system are given by
\be
\label{ThermoVars}
S_{BH} = \frac{2\pi r_+}{4G_3}
~~,~~~
T_H = \frac{r_+^2-r_-^2}{8 G_3 \ell^2 r_+}
~~,~~~
\Omega = \frac{r_-}{\ell r_+} ~.
\ee
The ensemble geometry should reflect generic features of the typical microstate, under the assumption that a generic microstate gives generic answers to sufficiently coarse-grained observables.
The inner horizon plays a prominent role in the thermodynamics.  This suggests that it ought to be a prominent feature of the microstates that are being averaged over in the ensemble geometry.  

A useful feature of the three charge system is that the canonical extremal geometry has a large smooth horizon, and already a large entropy in the BPS limit.  This contrasts with other systems like N=4 SYM or matrix theory, where the BPS limit fights with strong curvature because features of the horizon become of stringy or Planckian dimensions.  In the three charge system, one has a big sphere or ring in the BPS limit whose area counts a macroscopic entropy of BPS microstates. 

If the microstate geometries program were to be maximally successful, the density of states might be made of semiclassical geometries (though it might take a collection of semiclassical states of strings, branes \etc. in addition to geometry to fully enumerate the microstates).  Furthermore, the geometry is stationary and BPS, and so one might expect to be able to count states in the bulk theory using nonrenormalization theorems and localization techniques; such an approach has been spectacularly successful in certain model situations%
~\cite{%
Denef:2002ru,%
Maoz:2005nk,%
Rychkov:2005ji,%
Grant:2005qc,%
Denef:2007vg,%
Balasubramanian:2008da,%
deBoer:2008fk,%
deBoer:2008zn,%
Bena:2012hf,%
Dabholkar:2012zz%
}

In section~\ref{sec:MicroGeom} below, we will review relevant aspects of the program to construct BPS microstate geometries.  By the connection between horizons and thermodynamics, each of these microstate geometries should be horizonless because they each represent individual contributions to the ensemble rather than the ensemble itself, or even a sub-ensemble.  An ensemble geometry such as~\pref{BTZgeom} is instead realized as the one-point function of the metric in the ensemble of microstates; it is not itself realized on any particular microstate.%
\footnote{An alternative argued in~\cite{Sen:2009bm} is that the states constructed so far are distinct from the states that contribute to the ensemble geometry, and that the two contributions should be added together.}  
Instead the constructed microstate geometries cap off in the vicinity of the would-be horizon of the extremal ensemble geometry.  This ideology has had some success in generating the properties of two-charge geometries (see%
~\cite{Lunin:2002bj,Mathur:2005zp,Mathur:2007sc,Chowdhury:2008uj,Mathur:2008nj,Skenderis:2008qn,Balasubramanian:2008da,Chen:2014loa}
for example) 
though it should be stressed that the interpretation of the ensemble geometry is often somewhat suspect due to strong curvature near the horizon (see for example~\cite{Sen:2009bm,Chen:2014loa}.


The inner horizon appears to be a special place in the black hole geometry, so let us explore its properties a bit further.  In the analytically continued stationary ensemble geometry, the inner horizon is another bifurcate surface where the norm of the Killing vector changes sign.
Even outside the outer horizon, its effects are felt as a subleading singularity in the wave equation for linearized perturbations.  In $AdS_3$, the inner horizon is detectable in the monodromy of the frame field and spin connection, arbitrarily far from the source.  Any attempt to excise the inter-horizon region due to a `firewall', or some sort of `complementarity' (fuzzball or otherwise), will have to come up with an explanation of all the thermodynamic properties encoded by the inner horizon.

Typical practice in quantum gravity is to construct the average one-point function of the geometry in the ensemble of states (\ie\ the {\it ensemble geometry}) using the effective action, perhaps with leading higher derivative and semiclassical corrections.%
\footnote{In supersymmetric situations, this has been raised to a high art, see~\cite{Dabholkar:2012zz} for a recent review.}
This geometry has an outer and inner horizon, and if we naively continue further, a timelike singularity.  

\begin{figure}
\centerline{\includegraphics[width=3.0in]{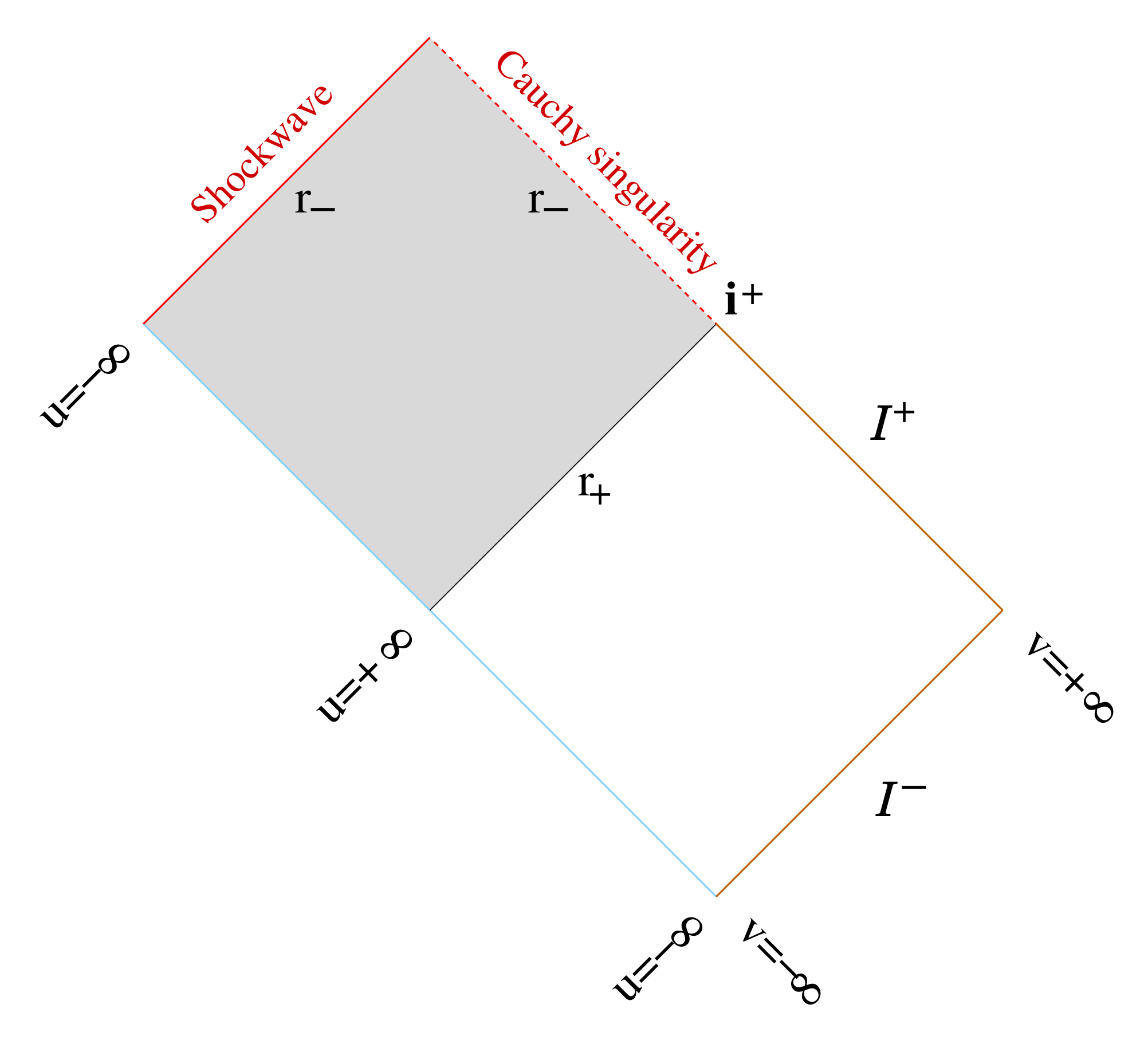}}
\setlength{\unitlength}{0.1\columnwidth}
\caption{\it 
Penrose diagram for charged/rotating black holes, taking into account the instability of the inner horizon, after%
~\cite{Marolf:2010nd,Marolf:2011dj}.  Instabilities of the analytically continued stationary solution preclude the existence of a regular geometry beyond the inner horizon.
}
\label{fig:instability}
\end{figure}

The inner horizon is however the locus of dynamical instabilities~\cite{Marolf:2010nd,Marolf:2011dj}.  In the analytically continued ensemble geometry, the inner horizon has ingoing and outgoing components, see figure%
~\ref{fig:instability}.  
The ingoing component is a Cauchy horizon, argued by%
~\cite{Hiscock1981110,Poisson:1990eh}
to be the locus of a weak null curvature singularity; in any event, black hole evaporation will replace this region by something else.  The outgoing component has been argued to be the location of a shock wave singularity.%
\footnote{Thus perhaps the true firewall is the null singularity at the inner horizon, rather than the one proposed at the outer horizon in%
~\cite{Almheiri:2012rt} (see also~\cite{Braunstein:2009v1}).}  
Thus, it seems likely that while the ensemble geometry can be analytically extended past the inner horizon, this part of the geometry is unstable and closes off.  What we seek in string theory is a mechanism to resolve this null singularity, along the lines of the many successes of string theory at resolving timelike singularities.  Typically the resolution of the latter is due to the appearance of new light degrees of freedom at the would-be singularity; it is just such a mechanism that we propose in this work.

The null singularity at the outgoing inner horizon, and Cauchy singularity at the ingoing inner horizon, lead to an excision of the regions beyond, leaving us with the Penrose diagram of figure~\ref{fig:instability}.  The shaded interior region beyond the outer horizon represents an analytic continuation of the exterior geometry.  For the purposes of the present discussion, we will treat the black hole inter-horizon region as physically relevant for revealing thermodynamic aspects of the black hole ensemble.

Support for this idea comes from the capped microstate geometries that have been constructed to date which all close off at or before reaching the horizon.  If capped geometries are the generic microstates of the ensemble of BPS states, then the ensemble average geometry will end in the vicinity of the horizon.  The question is, which horizon?  In the extremal case, the inner and outer horizons coincide.  It has long been suggested that the effects of the black hole interior can be modeled on a timelike stretched horizon slightly outside the outer horizon of the ensemble geometry%
~\cite{Thorne:1986iy}.  
One might think that the cap observed in the microstates constructed to date is a realization of the stretched outer horizon of the black hole -- that it is the membrane of the membrane paradigm.  However it's hard to tell since the inner and outer horizons coincide in the extremal case.

It is often stated that the `fuzzball' proposal for nonextremal black holes (see~\cite{Mathur:2008nj} for a review) resolves the singularity of the black hole at the {\it outer} horizon, and that in some sense the fuzzball is the membrane of the membrane paradigm.  One proposal%
~\cite{Mathur:2008kg} 
suggests the region inside the outer horizon of the geometry isn't accessed, that upon crossing the outer horizon one immediately tunnels into the ensemble of $e^{S_{BH}}$ fuzzball states of the black hole.  In this way of thinking,  the inner horizon (and all else inside the black hole) are only a convenient fiction -- virtual `dual' entities describing the vibrations of a `real' membrane at the outer horizon.  This notion has been dubbed `fuzzball complementarity'%
~\cite{Mathur:2012jk}.

It seems however that if the microstates program has any real power, it is that the classical solutions and the objects that populate them should have some relevance to the `ensemble geometry' that counts the states thermodynamically but has forgotten the microstructure.  For instance, the thermodynamic variables such as the locations $r_\pm$ of the inner and outer horizons, their surface gravities $\kappa_\pm$ and areas $A_\pm$, should be reflected in the structure and dynamics of typical microstates.  What might differ however is the global, long-time behavior of quantum correlations in the averaged, `ensemble geometry' as opposed to that of individual microstate geometries.

We propose here that in the nonextremal case, the cap of the microstate geometry is composed of `long string' degrees of freedom that oscillate, and that the inner and outer horizons represent the extremes of the long string's average motion.  In this way, the departure of the geometry from the ensemble average geometry remains small outside the outer horizon.  The region between the outer and inner horizons exists as a region of excitation of the `cap' of the microstate; but the geometry indeed does not exist beyond the inner horizon, as suggested by the analysis of%
~\cite{Marolf:2010nd,Marolf:2011dj} -- it is excised by the cap.  The extremely low tension of the long string suggests that the inter-horizon region will appear nearly indistinguishable from the vacuum to a freely falling observer. 

Some recent observations about black hole thermodynamics resonate with this picture.  A remarkable and mysterious role in this thermodynamics is played by the inner horizon of the analytically continued ensemble geometry.  Black hole thermodynamics relates properties of the outer horizon to those of the thermodynamics -- the outer horizon area in Planck units is the entropy, and the (Hawking) temperature is the surface gravity of the outer horizon
\be
\label{OuterThermo}
S_{BH} = S_+ = \frac{A_+}{4 G}
\quad,\qquad
T_H = T_+ = \frac{\kappa_+}{2\pi} ~.
\ee
These enter into a first law relation
\be
dM = T_+ dS_+ + \Omega^+ dJ + \Phi_e^+ dQ_e + \Phi_m^+ dQ_m ~,
\ee
where $\Omega^+$ is the angular velocity of the outer horizon, $J$ the angular momentum, $\Phi_{e,m}^+$ the electric/magnetic potentials, and $Q_{e,m}$ the electric/magnetic charges.

There is also a first law for the inner horizon:
\be
\label{InnerThermo}
dM = T_- dS_- + \Omega^- dJ + \Phi_e^- dQ_e + \Phi_m^- dQ_m ~.
\ee
For example, for BTZ black holes one has
\be
S_- = \frac{A_-}{4G} = \frac{2\pi r_-}{4G_3} 
~~,~~~
T_- = \frac{\kappa_-}{2\pi} = \frac{r_-^2-r_+^2}{8G_3 \ell^2 r_-}
~~,~~~
\Omega_- = \frac{r_+}{4G_3 r_-} ~;
\ee
these expressions are the same as in~\pref{ThermoVars}, with $r_+$ and $r_-$ interchanged.

The meaning of these inner horizon quantities from the gravitational point of view has to date been rather less clear than that of their outer horizon counterparts, which are physical observables apparent to exterior observers.  However, there is a remarkable relation which ties together the inner and outer horizon areas~\cite{Larsen:1997ge}:
\be
\label{areaprod}
\frac{S_+ S_-}{4\pi^2} =  f(q,J) \in \bZ ~,
\ee
where $f(q,J)$ is an integer valued function of the integer charges of the black hole.
This relation seems to be quite general -- it holds in all examples of four and five dimensional black holes and rings where it has been checked%
~\cite{Cvetic:2010mn,Castro:2012av,Cvetic:2013eda,Chow:2014cca},
though no proof is known.%
\footnote{In cases where the geometry has more than two horizons, a generalization holds involving the product over all the horizons, including complex ones.}

More explicitly, for asymptotically $AdS_3\times S^3$ black holes one has
\be
S_+ S_- = 4\pi^2 (q_1q_2q_3  + J_R^2 - J_L^2)
\ee
where $q_i$ are the number of integer quanta of each species of mutually BPS background charge (for instance D1-D5-P), and $J_{L,R}$ are the $\bS^3$ angular momenta (R-charges).

It seems reasonable to regard~\pref{areaprod} as a sort of rigidity property of the black hole interior -- the product of the areas of the inner and outer horizons is independent of the black hole mass, as well as the moduli or any other geometric data; it only depends on the quantized charges.
In particular, if one excites an $AdS_3$ black hole, the outer horizon will move further out and the inner horizon further in, but the geometric mean of the horizon radii 
\be
S_+ S_-  = \frac{\pi^2 r_+ r_-}{4 G_3^2} = 4\pi^2 q_1q_5 J_3
\ee
will stay fixed.

This fact about horizons is the sort of property one might expect if the microstate geometries were all capped, and the effect of adding energy was to (further) excite the cap.  Starting with the extremal geometry, where the cap is stationary at a fixed radius $r_+=r_- \equiv r_\ext$, adding energy should vibrate the cap like a membrane, at least for small excitations, and it is tempting to associate the expectation values of the minimum and maximum radial extent of the cap degrees of freedom with $r_\pm$.  The cap motion takes place about an `equilibrium' position which is the extremal radius $r_{\ext}$ for those charges, though as argued above the bulk of the cap degrees of freedom are located at the inner horizon for modest excursions from extremality.
A small disturbance of the cap is what is seen for small excitations of the smooth extremal microstate geometries constructed to date.  Such excitations have been considered in%
~\cite{%
Bena:2008nh,%
Bena:2008dw%
}
for probes that are mutually BPS with the background (and so change the background charges), and%
~\cite{%
Jejjala:2005yu,%
Giusto:2007tt,%
Bena:2011fc,%
Bobev:2011kk,%
Vasilakis:2011ki,%
Bena:2011ca,%
Bena:2012zi,%
Bena:2013gma%
}
for non-BPS excitations.

This picture of nonextremality differs from the standard `membrane paradigm' for black holes%
~\cite{Thorne:1986iy}.  
There, the membrane is effectively a phenomenological boundary condition somewhat outside the outer horizon, and a set of thermodynamic responses, that encode how the reservoir of black hole interior states interacts with its environment. Here instead, the cap extends over the black hole interior and represents a qualitative characterization of those interior degrees of freedom.

For $AdS_3\times \bS^3$ black holes, sums and differences of inner and outer horizon thermodynamic variables define quantities that are natural from the perspective of the dual CFT%
~\cite{Cvetic:1996xz,Cvetic:1996kv,Larsen:1997ge,Cvetic:1997uw,Cvetic:1997xv,Cvetic:1997vp,Cvetic:1998xh, Cvetic:1999ja}.
For instance,
\be
S_{L,R} = \half(S_+\pm S_-)
\quad,\qquad
\frac{1}{T_{L,R}} = \frac{2\pi}{\kappa_+} \pm \frac{2\pi}{\kappa_-}
\ee
are the entropies and temperatures of the left-moving and right-moving degrees of freedom of the dual CFT; and the two angular momenta $J_{L,R}$ on $\bS^3$ are naturally associated to the corresponding CFT chirality.  We now ask what thermodynamics tells us about where these left- and right-movers are located in a typical black hole state.


\subsection{Hints from the covariant entropy bound}
\label{sec:EntropyBound}

The question thus arises, what is the interpretation of these left- and right-moving quantities from the gravitational perspective, which are so natural from the point of view of the dual CFT?  A hint comes from thinking about an adiabatic interpolation between the two horizons.  The covariant entropy bound%
~\cite{Bousso:1999cb,%
Bousso:1999xy,%
Flanagan:1999jp,%
Bousso:2014sda}
states that the entropy that crosses a light sheet is bounded by the change in the area of the light sheet
\be
\label{BoussoBd}
\Delta S \le \frac{\Delta A}{4 G} ~.
\ee
Black holes are supposed to saturate this bound.  

Ordinarily, the bound~\pref{BoussoBd} is applied to processes where objects are thrown into a black hole, and the outer horizon area increases as a result of the stress-energy crossing the horizon.  That stress-energy is associated to an entropy through the equation of state of the matter, which obeys the bound~\pref{BoussoBd}; ordinary matter doesn't come close to saturating the bound, but one can approach it by adiabatically lowering another black hole through the horizon of the one under consideration.

\begin{figure}
\centerline{\includegraphics[width=5in]{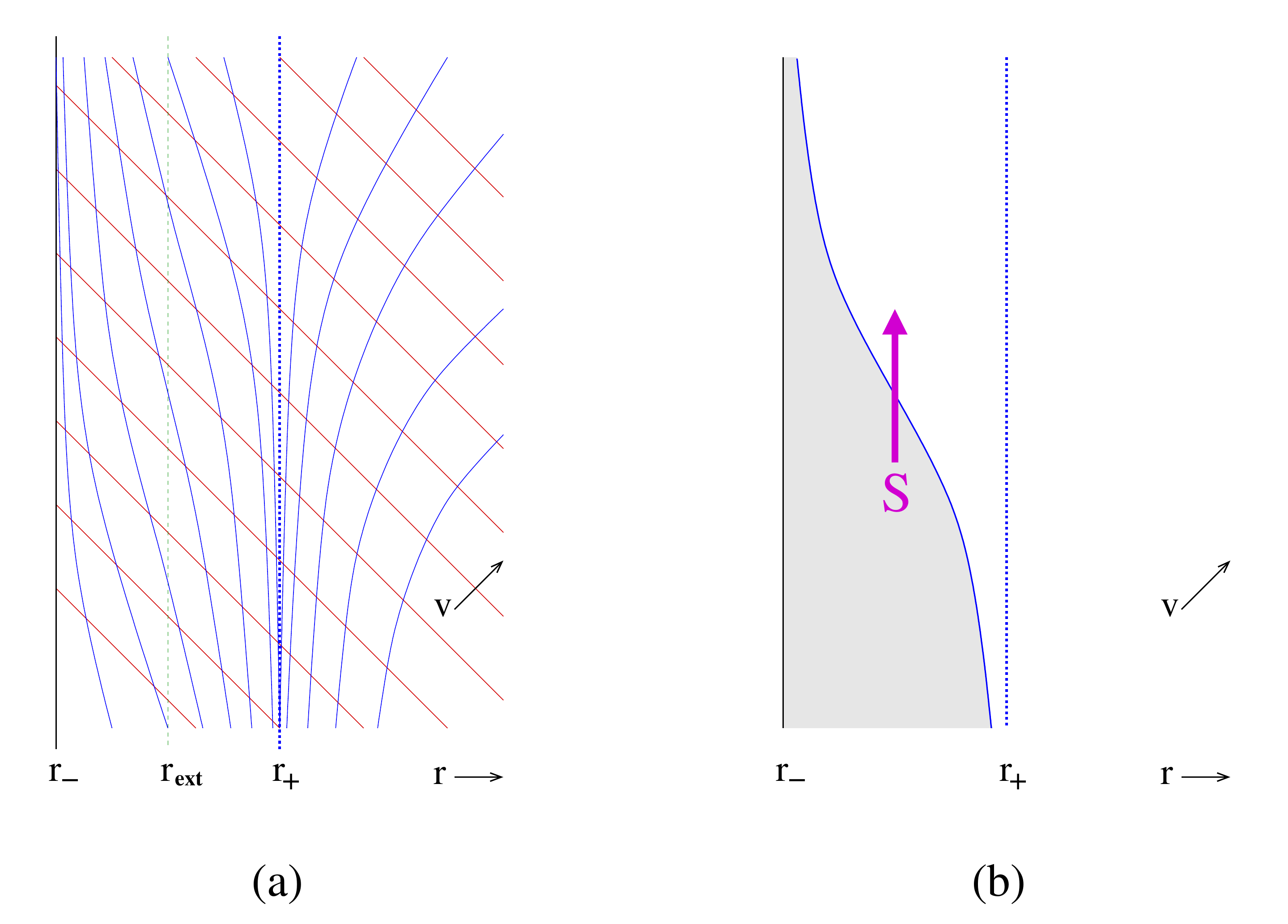}}
\setlength{\unitlength}{0.1\columnwidth}
\caption{\it 
(a) Near-horizon causal structure of Reissner-Nordstrom and BTZ geometries in Eddington-Finkelstein coordinates.  Outward going null trajectories are depicted in blue, ingoing in red.  The radius of the horizon of the extremal geometry carrying the same conserved charges ($r_+r_-=r_\ext^2$ in the BTZ case) is the green dashed line. 
(b) The covariant entropy bound applied to an outgoing light sheet that begins just inside the outer horizon and ends just outside the inner horizon.  The interior of the light sheet is shaded, and the flow of entropy across it indicated.
}
\label{fig:EFdiagrams}
\end{figure}

Here we wish to consider a rather different application of the bound~\pref{BoussoBd}, to the interior of the stationary black hole rather than to properties of the outer horizon under perturbations.
It will prove convenient to cast the geometry in Eddington-Finkelstein or Kerr coordinates.  
The Eddington-Finkelstein diagram is perhaps a bit more conducive to intuition than the Penrose diagram.  The latter is almost certainly misleading when it comes to the causal propagation of information from the interior to the exterior of the black hole, so we might as well make the horizons run approximately vertically until we have an appropriate notion to replace classical causal structure.
In the BTZ geometry, upon substituting
\be
dv = dt + \frac{dr}{N^2}
\quad,\qquad
d\varphi = d\phi - \frac{N_\varphi}{N^2} dr
\ee
the metric becomes
\be
\label{EFBTZ}
ds^2 = -N^2 dv^2 + 2dv \,dr + r^2(d\varphi+N_\varphi dv)^2  ~.
\ee
The inward and outward going null trajectories are depicted in figure~\ref{fig:EFdiagrams}a.

Consider an outward directed light sheet initially just inside the outer horizon at $r=r_+$, see figure%
~\ref{fig:EFdiagrams}b.  
Slowly the light sheet drifts inward until it asymptotes to the inner horizon at $r=r_-$.  In the process, its area changes from the area of the outer horizon to the area of the inner horizon; the change in area is 
\be
\label{TwoSR}
\frac{\Delta A}{4G} = \frac{A_- - A_+}{4G} = -2S_R
\ee
This clearly suggests that one should associate the `right-moving' entropy of black holes to degrees of freedom in the region between the two horizons, since these degrees of freedom will have crossed the light sheet during the course of its traverse of this region.  The sign is negative because the change is interpreted as the microstate degrees of freedom being transported out of the interior of the light-sheet as it moves inward.  Similarly, when not too far from extremality, the majority of the `left-moving' degrees of freedom comprise the contribution to the total entropy from degrees of freedom not traversed by the light sheet -- in other words, the constituents of the inner horizon, which are expected to resolve the singularity there.  One has
\be
S_L = S_+ - S_R = S_- + S_R
\ee
It is natural to conjecture that the inner horizon is the location of (most of) the cap, to the extent that it can be localized; and that in exciting the black hole above extremality, this is where most of the microstate degrees of freedom responsible for the entropy of the BPS spectrum have migrated to.  The covariant entropy bound is trying to tell us where the degrees of freedom carrying the black hole entropy are on average located; and that some of these degrees of freedom are allowed to `float' in the interior of the black hole between the two horizons, and are not forced to fall into the singularity as ordinary matter must.

\begin{figure}
\centerline{\includegraphics[width=5in]{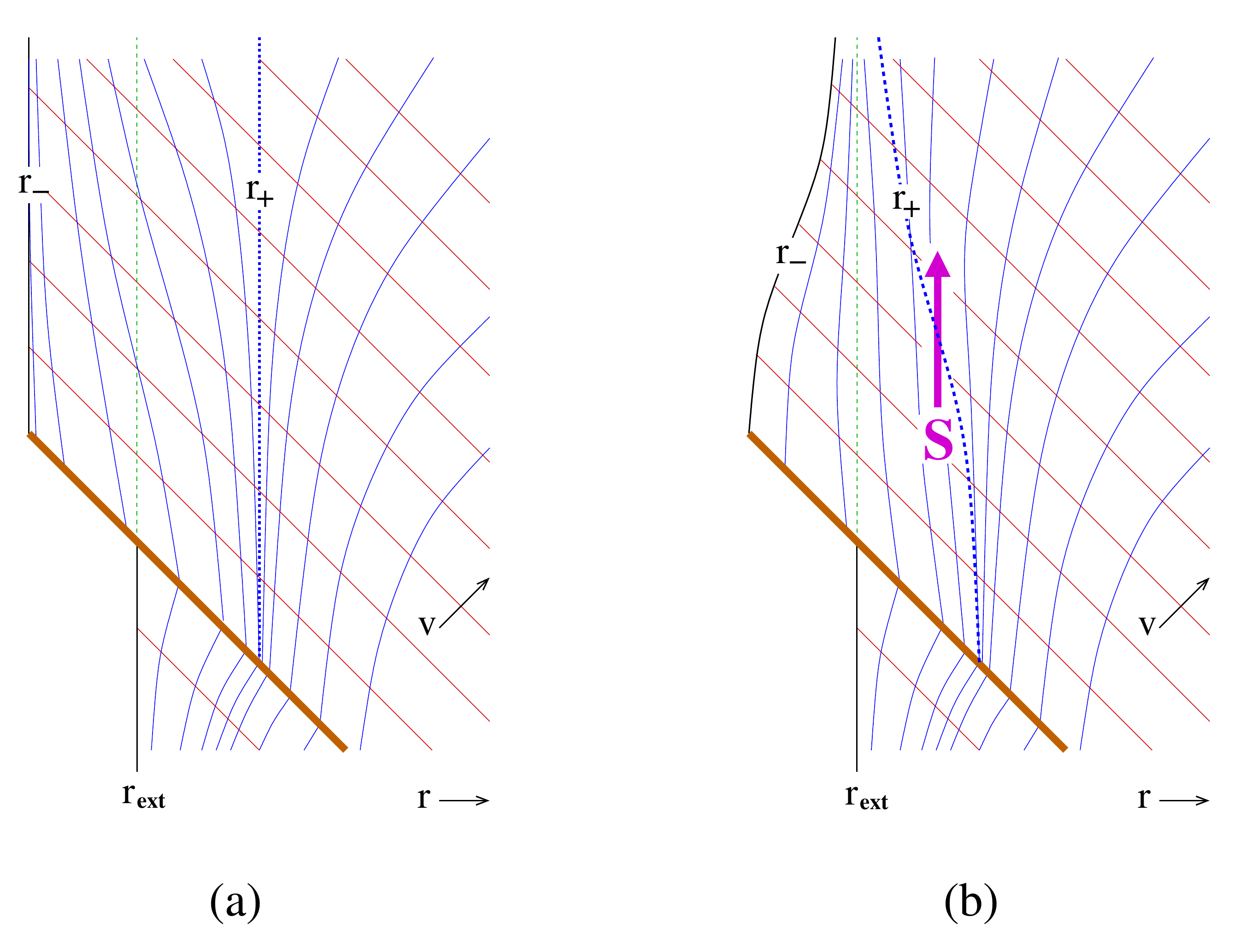}}
\setlength{\unitlength}{0.1\columnwidth}
\caption{\it 
(a) Geometry of a shockwave excitation (the brown null trajectory) of the extremal geometry. 
(b) Qualitative picture of the evaporation process when quantum effects are included.
}
\label{fig:shockwave}
\end{figure}

The factor of two in~\pref{TwoSR} prevents an interpretation of the inter-horizon degrees of freedom as consisting only of right-movers.  Qualitatively, one might think of the situation as follows.  At extremality, the macroscopic degeneracy of states resides in a set of cap degrees of freedom at the horizon.  The inner and outer horizons coincide, so it's ambiguous which horizon they should be associated with.
When the black hole is excited above extremality, the two horizons `delaminate'.  
A depiction of the splitting apart of the two horizons in response to an ingoing null shock is depicted in figure~\ref{fig:shockwave}a (in the classical theory; a cartoon of the evaporation process is depicted in figure~\ref{fig:shockwave}b).

Upon excitation, the inner horizon moves in and the outer horizon moves out; near extremality one has
\be 
r_\pm = r_\ext\pm \delta+\OO(\delta^2) ~,
\ee
and roughly half of $\Delta S$ in~\pref{BoussoBd} comes from the outer horizon moving out, and the other half from the inner horizon moving in.  We should associate half of this process to exciting the right-movers, and the other half to further exciting the left-movers; and that while the bulk of the cap is located at the inner horizon, there are some of its original degrees of freedom in the region between the inner and outer horizon.  These left-moving degrees of freedom are needed near the outer horizon to combine with the right-movers and emerge as Hawking radiation, since emitted quanta carry both left and right  conformal dimensions.

It is natural to conjecture that these $2S_R$ degrees of freedom in the inter-horizon region, that flow across the light-sheet of figure~\ref{fig:shockwave}b, are the `left- and right-movers' that are `available' to emerge as Hawking radiation.  
Especially for small nonextremality $(r_+-r_-)/r_+\ll 1$, the bulk of the entropy sits on the inner horizon, which one might regard as the actual firewall of the black hole (note that for large charges, the inter-horizon region can be as macroscopic and weakly curved as desired).  By treating the covariant entropy bound this literally, our major remaining task will be to explain how the degrees of freedom carrying this entropy are not forced to fall into the singularity as the causal structure felt by ordinary matter would dictate, but rather seem to float in the region between the two horizons.  A proposed explanation will be given in the final section below.

The inner and outer horizons have a surface gravity $\kappa_\pm$ and associated temperatures $T_\pm$ via~\pref{OuterThermo},\pref{InnerThermo}.  It seems reasonable to associate a `local temperature' to the region in between, especially near extremality where the temperature is small, however because the outgoing light-sheet is not static in the inter-horizon region, such a concept is necessarily somewhat ambiguous.  For spherically symmetric spacetimes, a proposal for a definition of surface gravity was given in%
~\cite{Fodor:1996rf,Nielsen:2007ac}:
\be
\kappa = - n^\mu l^\nu\nabla_\nu l_\mu
\ee
where $l^\mu$ and $n^\mu$ are outgoing and ingoing null normals to a sphere at fixed $r$, satisfying $l^\mu n_\mu = -1$; and furthermore, $n^\mu$ is chosen to be the tangent vector to an affinely parametrized ingoing null geodesic whose affine parameter is normalized at spatial infinity by $t^\mu n_\mu=-1$ in terms of the asymptotic timelike Killing vector $t^\mu$.  This definition agrees with the usual notion of surface gravity on a bifurcate Killing horizon, and smoothly extends it off that surface using ingoing null trajectories (the surfaces of constant $v$ in Eddington-Finkelstein coordinates).  Evaluated on the BTZ metric~\pref{EFBTZ}, one finds
\be
\label{LocalTemp}
\kappa(r) = \frac{ -3 r^4+r^2(r_+^2+r_-^2)+r_+^2 r_-^2}%
{16 G_3\ell^2 r \bigl[r^4 - r^2(\ell^2 + r_+^2 + r_-^2)+r_+^2r_-^2\bigr]} ~.
\ee
This quantity is naturally positive at the outer horizon where the outgoing null geodesics are diverging, negative at the inner horizon where they are converging, and smoothly interpolating in between.

It may seem odd that the temperature of the inner horizon is negative.  In thermodynamic terms, this simply means that the entropy of this subsystem {\it decreases} as one adds energy to the system.  For the black hole, this means that the inner horizon `cap' carries less and less of the total entropy of the black hole as it is further excited; more and more of the entropy is instead carried in the inter-horizon region, until one approximates a Schwarzschild black hole and the inner horizon carries essentially no entropy.  In this sense, the inner horizon represents a reservoir of degrees of freedom that is tapped to fill the inter-horizon region when the black hole is excited above extremality.

In what follows, we will build a picture of the black hole interior as the dominant support of the wavefunction of the `long string' that holds its entropy.  We will interpret the microstate geometries program as giving hints about the nature and location of this long string, near but just below the black hole transition.

In a resolution of the information paradox, there are two places where magic has to happen: First, at the inner horizon, something has to resolve the singularity, and store incoming information in its degrees of freedom; this is what the cap does in the extremal microstate geometries constructed to date, assuming they are stable under small perturbations.  Second, something nonlocal has to allow information -- now trapped on the resolved null singularity in the Eddington-Finkelstein diagram -- to cross over from the black hole interior to its exterior.  A longstanding idea of how this might happen uses the fuzziness of light cones in string theory%
~\cite{%
Martinec:1993jq,%
Lowe:1994ns,%
Lowe:1995ac%
}
to try to pass information outside the light cones of the effective geometry, but how precisely this could work and how it could operate on the necessary macroscopic scales was never made clear.  Our proposal here uses the fractionated tension of the long string sector to extend the nonlocality of ordinary strings over the $AdS$ scale and beyond, such that the fuzziness of the long string's light cones does not resolve the distinction between the inner and outer horizon~-- that the reason there are degrees of freedom that seem to float in the inter-horizon region is that they are not subject to the light cone structure felt by ordinary matter.


\section{Review of supergravity microstate solutions}
\label{sec:MicroGeom}

We now turn to a review of the basics of the construction of BPS microstate geometries, and how they cap off the geometry near the would-be horizon, following%
~\cite{Warner:2008ma,%
Bena:2008dw,%
Niehoff:2013kia,%
Gibbons:2013tqa%
}.  
This discussion will be followed in section~\ref{sec:Quivers} by a summary of the associated quiver quantum mechanics construction of these states via the quantization of the collective coordinates of their underlying brane constituents, following%
~\cite{%
Denef:2002ru,%
Denef:2007vg,%
deBoer:2008fk,%
deBoer:2008zn,%
Manschot:2010qz,%
Lee:2011ph,%
Bena:2012hf%
}.  
This overview will lay the foundation for a discussion of the limits of validity of supergravity in these solutions, and eventually a physical picture will emerge of the mechanism underlying the breakdown of the supergravity description that will be the focus of section~\ref{sec:Fivebranes}.


\subsection{BPS geometry}

There are two useful duality frames in which to consider the three-charge systems of interest.  The first is M-theory compactified to 5d on $\bT^6$, where the charges and dipole charges are
\be
\label{Mcharge}
\begin{tabular}{| l | l |}
\hline 
Conserved Charge 
&
Dipole Charge
\\
\hline
M2:~$5\,6$
&
M5:~ $7\,8\,9\,10\,\psi$
\\
\hline
M2:~$7\,8$
&
M5:~ $5\,6\,9\,10\,\psi$
\\
\hline
M2:~$9\,10$
&
M5:~ $5\,6\,7\,8\,\psi$ 
\\
\hline
\end{tabular}
\ee
Here $5...10$ are the torus directions, and $\psi$ is an angular coordinate along which the dipole charge is distributed.  The symmetric arrangement of brane sources simplifies a variety of expressions for the supergravity fields.

Shrinking the $\bT^2$ of the 9-10 directions takes us to type IIB string theory on a dual circle $\tilde 9$, with the charges and dipoles
\be
\label{IIBcharge}
\begin{tabular}{| l | l |}
\hline 
Conserved Charge 
&
Dipole Charge
\\
\hline
D3:~$5\,6\, \tilde 9$
&
D3:~ $7\,8\,\psi$
\\
\hline
D3:~$7\,8\,\tilde 9$
&
D3:~ $5\,6\,\psi$
\\
\hline
P:~$\tilde 9$
&
KK:~ $5\,6\,7\,8\,\tilde 9\,\psi$ 
\\
\hline
\end{tabular}
\ee
where in the last dipole, the $\tilde 9$ circle is nontrivially fibered in the KK monopole.  The D1-D5 duality frame is simply related to this one by T-duality.
The duality holds for all three-charge solutions, however only when the fields are constant along the circles being dualized will there be a simple relation between their explicit forms in the two duality frames.  


The duality frame of particular interest to us is the D1-D5-P frame, where one can take an $AdS_3 \times\bS^3$ scaling limit.  
The metric takes the form
\be
\label{6dMetric}
ds^2 = -\frac{2}{\sqrt{Z_1Z_2}} \bigl(dv+\beta\bigr)\bigl(du+k+ \hf \FF (dv+\beta)\bigr) + \sqrt{Z_1Z_2}\, ds_4^2(\BB)
\ee
The BPS equations that determine the coefficient functions in this metric, together with the other supergravity fields, have been extensively studied, and have a remarkable linear structure, allowing for explicit solutions to be constructed.  Supersymmetry implies that these functions are independent of the time coordinate $u$.


The equations simplify dramatically~\cite{Bena:2005va} if one in addition assumes that the solutions are independent of $v$; this also simplifies the duality relation to the M-theory frame.
A further simplification assumes that the base $\BB$ has a tri-holomorphic Killing vector isometry, \ie\ that it is a Gibbons-Hawking space.  In this circumstance, the metric is written as
\be
\label{6dmetric}
ds_4^2 = V^{-1}(d\psi+A)^2 + V d\vec y\cdot d\vec y \quad,\qquad \vec\nabla V = \vec\nabla\times\vec A 
\ee
with $V$ harmonic on the flat $\bR^3$ parametrized by $\vec y$.
In terms of the frame forms
\bea
\Omega_\pm^{(i)} &=& e^0\wedge \ehat^i \pm {\textstyle\half} \epsilon_{ijk} \ehat^j\wedge \ehat^k 
\nonumber\\
\ehat^0 &=& V^{-1/2} (d\psi+A) \quad,\qquad  { \ehat^i}= V^{1/2} d{y^i} ~,
\eea
the self-dual two-forms
\be
\Theta^{(I)} \equiv \Omega_+^{i} \partial_i(V^{-1} K^I)
\ee
are closed (therefore co-closed and harmonic) provided $K^I$ is harmonic.  
The vector potential $\beta$ can be expressed as
\be
\beta = \frac{K_3}{V}\bigl(d\psi+A\bigr) + \vec\xi\cdot d\vec y
\quad,\qquad
\vec\nabla \times \vec \xi = -\vec\nabla K_3 ~.
\ee

Suitable choices are
\be
V = \epsilon_0 + \sum_{a=1}^N \frac{q_a}{r_a} \quad, \qquad K^I = \kappa_0^I +  \sum_{a=1}^N \frac{k_a^I}{r_a}
\ee
with $r_a=|\vec y - \vec y_a|$ the distances from sources in the various harmonic functions.  
One demands $q_0=\sum_a q_a =1$ so that the four-manifold is asymptotically $\bR^4$.  When $|q_a|=1$, the base $\BB$ is locally $\bR^4$ near the source location $\yy_a$;  sources with $|q_a|\ne1$ have a $\bZ_{q_a}$ orbifold singularity near the source.
The fiber coordinate $\psi$ of the Gibbons-Hawking geometry degenerates at the poles of $V$, determining a two-cycle $\Delta_{ab}$ consisting of this fiber circle times any path between $\vec y_a$ and $\vec y_b$; the flux of the two-form $\Theta^{(I)}$ through this two-cycle is given by
\be
\label{Piflux}
\Pi^{(I)}_{ab} = \left(\frac{k^I_b}{q_b} - \frac{k^I_a}{q_a} \right) \quad,\qquad 1\le a,b\le N ~.
\ee

Defining $Z_3=-\FF$, the warp factors in the metric are then determined as 
\bea
Z_I &=& {\textstyle\half} C_{IJK} V^{-1}K^J K^K + L_I
\nonumber\\
k &=& \mu(d\psi+A) + \omega
\nonumber\\
\mu &=& \coeff16 C_{IJK}\frac{K^I K^J K^K}{V^2} + \frac{1}{2V} K^I L_I + M
\nonumber\\
\vec\nabla\times\vec\omega &=& V\vec\nabla M - M\vec\nabla V + \coeff12(K^I\vec\nabla L_I - L_I \vec\nabla K^I)
\eea
with
\bea
\label{KLdefs}
L_I = \ell_{0,I} + \sum_{a=1}^N \frac{\ell^a_I}{r_a}
\quad &,& \qquad
\ell^a_I = -\coeff12 C_{IJK} \frac{k^J_a k^K_a}{q_a}
\nonumber\\
M = m_0 + \sum_{a=1}^N \frac{m_a}{r_a}
\quad &,& \qquad
m_a = \coeff1{12} C_{IJK} \frac{k^I_a k^J_a k^K_a}{q_a^2} = \coeff12\frac{k^1_a k^2_a k^3_a}{q_a^2}  ~.
\eea
The $AdS_3\times \bS^3$ asymptotic form of the metric is achieved for
\be
\label{netcharges}
\epsilon_0= 0
~~,~~~
q_0 = \sum_{a=1}^N q_a = 1
~~,~~~
\kappa_0^I = 0 
~~,~~~
\ell_{0,I} = 0 
~~,~~~
m_0 = -\coeff12 q_0^{-1} \sum_{a=1}^N \sum_{I=1}^3 k_a^I ~.
\ee
Finally, the absence of closed timelike curves imposes the {\it bubble equations}
\be
\label{BubbleEqs}
\sum_{b=1,b\ne a}^N \frac{\langle\Gamma_a ,\Gamma_b\rangle}{r_{ab}} =  
\langle \Phi,\Gamma_a \rangle ~,
\ee
where $\Gamma_a$ is the eight-vector of charges, and $\Phi$ the harmonic potential background
\be
\Gamma_a = (q_a, \ell_I^a, k_a^I, m_a)
\quad,\qquad
\Phi \equiv (\epsilon_0,\ell_I^0,\kappa_0^I,m_0) = (0,0,0,m_0) ~;
\ee
the symplectic inner product $\langle*,*\rangle$ is
\be
{\langle\Gamma_a ,\Gamma_b\rangle} = 2(q_b m_a - q_b m_a) + \sum_{I=1}^3 (\ell_I^b k_a^I - k_b^I\ell^a_I ) ~.
\ee
Using~\pref{Piflux},\pref{KLdefs},\pref{netcharges} these conditions can also be expressed as
\be
\label{BubbleEqs1}
\sum_{b=1,b\ne a}^N \Pi_{ab}^{(1)}  \Pi_{ab}^{(2)}  \Pi_{ab}^{(3)} \frac{q_a q_b}{r_{ab}} = -2 m_0 q_a - \sum_{I=1}^3 k_a^I ~.
\ee
The bubble equations place $N-1$ constraints on the $3(N-1)$ parameters $\vec y_a$ (modulo translations) and the $4N-1$ parameters $q_a$, $k_a^I$.  

The conserved charges of the solution are given by
\bea
\label{BkgdCharges}
Q_I &=& -2 C_{IJK} \sum_{a=1}^N q_a^{-1} \tilde k^J_a \tilde k^K_a ~,
\nonumber\\
J_R &=& J_1+J_2 =\coeff43 C_{IJK} \sum_{a=1}^N q_a^{-2} \tilde k^I_a \tilde k^J_a \tilde k^K_a 
\\
J_L  &=& J_1-J_2 = 8|\vec D| \quad,
\nonumber
\eea
where
\be
\tilde k^I_a \equiv k^I_a - q_a \sum_{b=1}^N k^I_b 
\qquad,\qquad
{\vec D} \equiv \sum_{I} \sum_{a=1}^N {\tilde k}^I_a {\vec y}_a ~.
\ee
The conserved D1-D5-P background charges $Q_I$ of the solution, as well as the angular momenta $J_{L,R}$, are determined by the residues of the poles in the harmonic functions.  Thus each pole is the locus of some portion of the sources of background charge.
Another convenient way to think of the angular momentum $J_L$~\cite{Berglund:2005vb,Bena:2006kb}
distributes it among pairs of poles:
\bea
\label{Jpair}
\vec J_L &=& \sum_{a\ne b} \vec J^{ab}_{L}
\\
\vec J^{ab}_{L} &=& -8 \langle \Gamma_a,\Gamma_b\rangle\;\hat \yy_{ab}
\quad,\qquad \hat \yy_{ab} \equiv \frac{\yy_{ab}}{|\yy_{ab}|}
\nonumber
\eea

The assumptions that the metric is independent of $v$ and $\psi$ has simplified the system of BPS equations sufficiently that a reasonably explicit solution can be found, whose data consists of the locations and residues of the poles in the various harmonic functions, modulo various constraints.  These assumptions also allow the solution to be carried over to the dual M-theory background by simply copying over the corresponding harmonic functions.
However, the generic D1-D5-P BPS state will be both $v$ dependent (since the generic momentum excitation is $v$ dependent) and $\psi$ dependent (since the generic angular momentum excitation is $\psi$ dependent).  A general strategy for generating $(v,\psi)$ dependent solutions was outlined in~\cite{Niehoff:2013kia}.


\subsection{Solutions with less than three poles}

As a somewhat trivial example, consider only a single pole $V=1/r$ with $Z_{1,2}=L_{1,2} = n_{1,2}/4r$; $Z_3 = -\FF=0$, and all the dipole charges $k^I$ vanishing.  Then the metric~\pref{6dmetric} is simply $AdS_3\times \bS^3$ in Poincar\'e coordinates, or equivalently the extremal $M_3=J_3=0$ BTZ black hole.

If we generalize slightly to allow angular momentum, but still suppressing the dipole charges:
\be
V=\frac1r
~~,~~~
K^I = 0
~~,~~~
Z_I = L_I = 1 + \frac{Q_I}{4r}
~~,~~~
\mu = M = \frac{J}{8r}
~~,
\ee
the geometry describes (the U-dual of) a BMPV black hole%
~\cite{Breckenridge:1996is} (see also~\cite{Bena:2008dw}).  
There is a horizon at $r=0$ whose area yields the black hole entropy
\be
S_{BMPV} = 2\pi\sqrt{Q_1Q_2Q_3 - J^2}~.
\ee

The generalization to include dipole charges, still with a single pole, leads to black ring solutions (see%
~\cite{Bena:2008dw} again for a discussion in the present framework).
\be
V = \frac1r
~~,~~~
K^I = \frac{-q_I}{2|\yy-\yy_0|}
~~,~~~
L_I = 1 + \frac{Q_I + C_{IJK} q^J q^K}{4|\yy-\yy_0|}
~~,~~~
M = \frac{J}{16R} -\frac{J}{16|\yy-\yy_0|} 
~~,
\ee
where $R$ is the ring radius.  The conserved and dipole charges of the solution are $Q_I$ and $q_I$, respectively, and the angular momentum and entropy are
\be
J = 4(q_1+q_2+q_3)R
\quad,\qquad
S = \pi\sqrt{\II_4}
\ee
with $\II_4$ the quartic invariant
\be
\II_4 = (2 q_1q_2Q_1Q_2 - q_3^2Q_3^2 + {\rm cyclic}) - 4q_1q_2q_3 J  ~.
\ee

These single pole solutions are what we have termed ensemble geometries, in that they all have horizons at the pole of the harmonic functions; microstate geometries, on the other hand, should be everywhere smooth.  This is what is accomplished by the choices~\pref{KLdefs} of $\ell^I_a$, $m_a$ as the residues of the poles in $L^I$ and $M$; these choices guarantee that the harmonic functions $Z_I$ and $\mu$ remain finite everywhere, and the solution is smooth.

Solutions with two charge centers are worked out in%
~\cite{%
Giusto:2004ip,%
Giusto:2004id,%
Giusto:2004kj,%
Berglund:2005vb%
}.
Following~\cite{Giusto:2012yz}, parametrize the harmonic functions as
\be
\label{TwoPoleV}
V = -\frac{s}{|\yy|} + \frac{s+1}{|\yy-\cc|}
\quad,\qquad
K_I = d_I\Bigl( \frac1{|\yy|} - \frac1{|\yy-\cc|}\Bigr) ~.
\ee
The constraints~\pref{BubbleEqs1} then determine
\be
|\cc| = \frac{1}{s^2(s+1)^2}\frac{d_1d_2d_3}{d_1+d_2+d_3} 
~.
\ee
The dipole charges $d_I$ are related to integer quanta via
\be
d_1 = \frac{g\lstr^2}{2 R} k_1
\quad,\qquad
d_2 = \frac{g\lstr^6}{2V_4 R} k_2
\quad,\qquad
d_3 = \frac{R}{2} k_3
\ee
and in turn the $k_I$ are related to the conserved background charges and angular momenta via
\bea
\label{TwoCtrCharges}
& &n_1 = \frac{k_2 k_3}{s(s+1)}
\quad,\qquad
 n_5 = \frac{k_3 k_1}{s(s+1)}
\quad,\qquad
n_p = \frac{k_1 k_2}{s(s+1)} 
\nonumber\\
& &J_L = (s+\hf) \frac{n_1 n_5}{k_3}
\quad,\qquad
J_R = \frac{n_1n_5}{2k_3}
\eea
From this one sees that when $s=k_3$ or $s=k_3+1$ the geometry is the spectral flow by two units of the Ramond vacuum states $\ket{0^{++}}_R$ and $\ket{0^{-+}}_R$, respectively; for other values~\cite{Giusto:2012yz} gave an interpretation of the geometry in terms of `fractional spectral flow'.
The integer data $(s,k_1,k_2,k_3)$ specify a solution; proper quantization of the charges and angular momenta is ensured if $s(s+1)$ and $n_1n_5$ are integer multiples of $k_3$.  

While these solutions carry all three charges and both angular momenta, they are not typical microstates in that the excitation gap is typically large -- the throat of the geometry is not deep in the regime where one trusts the geometry.  The CFT duals of BPS microstates are characterized by the twisted sectors of the symmetric product orbifold, which are in turn specified by a word in the symmetric group.  The words in the symmetric group which realize the above geometries were identified in%
~\cite{Giusto:2012yz}
to consist of $n_1n_5/k_3$ cycles of length $k_3$; the twist ground state for each cycle is then spectrally flowed by an amount proportional to $s$.  The excitation gap in the dual CFT is thus $k_3^{-1}$ in AdS units, rather than the $(n_1n_5)^{-1}$ one expects of a typical microstate in the black hole ensemble.  

If one takes $k_3\sim n_1n_5$ in order to have the right gap, then there will be high order orbifold singularities at the poles of the sphere in the Gibbons-Hawking base $\BB$.
The smoothness of the geometry was investigated in~\cite{Jejjala:2005yu,Giusto:2012yz}.
It turns out that for $(k_3,s,s+1)$ all mutually prime, the geometry is completely smooth and the metric is locally $AdS_3\times \bS^3$.  When a pair has a common divisor, there are orbifold singularities of the order of that divisor; the orbifold quotient can be as large as $\bZ_{k_3}$.  For instance, if we take $s(s+1)=k_3$ and $k_3=n_1n_5$, we have orbifold singularities of order $s$ and $s+1$ at the poles of the $\bS^2$, which are of order $\sqrt{n_1n_5}$. 

These geometries cannot belong to an ensemble with macroscopic horizon of BMPV type~\cite{Berglund:2005vb}. 
The BMPV black hole~\cite{Breckenridge:1996is} has horizon area
\be
\label{Sbmpv}
S_{BMPV} = 2\pi\sqrt{n_1n_5 n_p - J_L^2}
\ee
vanishes in the limit $J_L^2\to n_1n_5n_p$ which is implied by~\pref{TwoCtrCharges}.  This is not so surprising, because when $k_3=n_1n_5$, the state is the spectral flow of the extremal BTZ black hole geometry.  These states are not on the verge of becoming BMPV black holes (\ie\ don't have a sufficiently deep throat and small excitation gap) unless $k_3\sim n_1n_5$; in this limit one has macroscopic  $J_L$ but $J_R$ of order one.  


\subsection{Three or more poles}

We now turn to a review of solutions with three or more poles, following%
~\cite{%
Bena:2005va,%
Bena:2006kb,%
Bena:2007kg,%
Bena:2007qc,%
Denef:2007vg%
}.  
In this case, the positions of the poles in the solution are not fixed by the charges as in the two pole case.  The major new feature in this case is the existence of scaling solutions for which the bubble equations can be solved for a one parameter family of pole locations where a subset $\SS$ of the poles collapse to the same point $\yy_0$%
~\cite{%
Denef:2000nb,%
Bena:2006kb,%
Denef:2007vg,%
Bena:2007qc,%
Bates:2003vx%
} (for a recent pedagogical discussion, see~\cite{Gibbons:2013tqa}). The bubble equations~\pref{BubbleEqs1} are approximately solved by letting
\be
{r_{ab}} \sim \lambda\; (\Pi_{ab}^{(1)}  \Pi_{ab}^{(2)}  \Pi_{ab}^{(3)}{q_a q_b})
= \lambda \langle \Gamma_a, \Gamma_b \rangle \equiv \lambda\; \Gamma_{ab}
\ee
for $a,b\in\SS$; $\lambda\to 0$ is the scaling limit that pushes this collection of sources together.

Let $\epsilon$ be the characteristic distance in $\bR^3$ between poles in $\SS$, and let $\eta$ be the distance from the cluster center $\yy_0$ to the nearest pole in the complement of $\SS$.  For $\epsilon\ll |\yy-\yy_0| \ll\eta$, the various harmonic functions scale as $c/r$ where $c$ is the sum of contributions from $\SS$.  In particular, the six-dimensional metric is locally $AdS_3\times \bS^3$ with a curvature radius determined by the total charges carried by $\SS$.  Inside the radius $\epsilon$, the geometry caps off.

Thus, one can worry that in the scaling limit $\lambda\to 0$, the throat created by the scaling cluster becomes infinitely deep, and the microstate develops a horizon of finite area in contradiction with the fact that microstates by themselves have no entropy and therefore, according to the Bekenstein-Hawking relation, should not have a finite area horizon.

The analysis of scaling solutions in~\cite {Bena:2007qc} pointed out the close connection between the moduli space of Gibbons-Hawking centers and the moduli space of D-brane bound states in four-dimensional supergravity; the dynamics of these centers was analyzed in%
~\cite{%
Denef:2000nb,%
Denef:2002ru,%
Denef:2007vg%
}
using the quiver gauge theory of the D-brane open string description.%
\footnote{The `moduli space' of solutions is a convenient fiction; really it is an attempt to isolate the structure of the lightest degrees of freedom in a particular corner of the configuration space.  These degrees of freedom are not true moduli like the asymptotic shape of the compactification torus.  Due to the low dimensionality of the conformal boundary, the modes in question are normalizable and fluctuate; they generically have time dependence and must be path-integrated over.  These deformations are not moduli of the background that are fixed data of the spacetime conformal field theory; rather they are soft modes of a particular solution or set of solutions that one hopes to treat properly by methods of collective coordinate quantization.  }
The authors of~\cite {Bena:2007qc} speculated that quantization of the moduli space could prevent the formation of a horizon.

The quantization of the moduli space of three centers was performed quite explicitly in~\cite{deBoer:2008zn}.  
Generally, the space of classical solutions is endowed with a symplectic structure~\cite{Crnkovic:1986ex}, but extracting it from the geometry and the supergravity action is complicated.%
\footnote{This exercise has been carried out successfully for the two-charge D1-D5 backgrounds in~\cite{Rychkov:2005ji}, using the explicit construction of the metrics of Lunin and Mathur~\cite{Lunin:2002bj} via the map to the duality frame in which the charges are F1-P and then quantizing the resulting effective string modes.  Supersymmetry is expected to protect these modes as being the priveleged collective modes of the supergravity fields in the original duality frame that one wishes to quantize.}
The quiver gauge theory description supplies a route to determining the symplectic form, and a nonrenormalization theorem supports the notion that the ground states of the quiver should match those of supergravity, and thus for the BPS states one should find the same symplectic form from the space of BPS supergravity solutions.  We turn now to an overview of the analysis of~\cite{deBoer:2008zn}. 


\section{4d black holes and Quiver QM}
\label{sec:Quivers}

Quantization of the collective coordinates of D-brane bound states has provided a great deal of insight into the BPS black hole spectrum%
~\cite{%
Denef:2002ru,%
Denef:2007vg,%
deBoer:2008fk,%
deBoer:2008zn,%
Manschot:2010qz,%
Lee:2011ph,%
Bena:2012hf%
}, 
see~\cite{DeBoer:2008zz} for a review.
In this section we summarize these results, which will prepare the way for a discussion of singularities in the following section.

\subsection{4d BPS solutions and their 5d M-theory uplift}

The 4d geometries sourced by D-brane charges have a description very similar to the 6d type IIB geometries we have been discussing.  An elegant analysis of Denef and collaborators%
~\cite{Denef:2000nb, Denef:2002ru, Bates:2003vx, Denef:2007vg}
constructs the near-horizon geometries and relates them to a variety of phenomena such as walls of marginal stability, \etc.
Much of the near-horizon structure is captured by an effective quiver quantum mechanics for the adiabatic motion of the D-brane centers.

One starts with 4d type IIA string theory, for simplicity consider a torus compactification, with a collection of $N$ (D6,D4,D2,D0) charged sources located at points $\yy_a$ in their transverse $\bR^3$; let the charge of the $a^{\rm th}$ source be
\be
\Gamma_a =  (p_a^0, p_a^A, q^a_A, q_0^a)
\ee
where $A=1...b_2$ labels a basis of two-cycles on the torus.
As with the 6d microstates construction, these objects will source the geometry via a set of harmonic functions
\bea
H^0 = \sum_a \frac{p_{a}^0}{ r_a} + h^0
\quad&,&\qquad
H_0 = \sum_a \frac{q^a_0}{ r_a}  + h_0
\\
H^A = \sum_a \frac{p_a^A}{ r_a} + h^A
\quad&,&\qquad
H_A = \sum_a \frac{q^a_A}{r_a} + h_A
\nonumber\\
\eea
(here $r_a=| \yy-\yy_a|$).  There is an overall integrability condition on the locations $\yy_a$ of the centers which plays the role of the `bubble equations'~\pref{BubbleEqs}
\be
\label{integrability}
\sum_b\frac{\langle \Gamma_a,\Gamma_b \rangle}{r_{ab}} = \langle h,\Gamma_a \rangle ~.
\ee
Here, $\langle *,*\rangle$ is again the symplectic product
\be
\langle \Gamma_{a}, \Gamma_{b} \rangle = -p_{a}^0 q_0^{b} + p_{a}^A q_A^{b} - q^{a}_A p^A_{b} + q_0^{a} p_{b}^0 ~,
\ee
and $r_{ab}$ is the inter-center separation.

Under suitable conditions, one can take an M-theory limit where an additional (fibered) circle appears in the geometry, 
and the brane charges (D6,D4,D2,D0) become (KK,M5,M2,P).  That is, D0 charge lifts to momentum along the M-theory circle; D2 branes are the membranes of M-theory; D4 branes are M5 branes wrapped around the extra circle; and D6 branes are KK monopoles in 11d, with the extra M-theory circle being the nontrivial fiber of the monopole solution.  As usual, one wants to decouple the branes from the ambient gravitational dynamics, in order that the near-source geometry is entirely captured by the quantum theory of open string degrees of freedom on the branes.  

In the M-theory limit, a charged 4d black hole naively becomes a 5d charged black ring smeared over the extra circle.  When the D6 charge is sourced by flux, the flux threads a two-sphere in the M-theory geometry, consisting of the M-theory circle fibered over a path between D6 charge centers, at which the circle pinches off.  The momentum along the fiber circle is now angular momentum, since the circle is contractible.
This picture connects 4d D-brane bound states to the 5d M-theory picture of section~\ref{sec:MicroGeom}.  For more details, see%
~\cite{Denef:2002ru,Balasubramanian:2006gi,deBoer:2008fk,deBoer:2008zn}.


In a decoupling limit, this type IIA geometry can be lifted to a five dimensional M-theory solution with $AdS_3\times \bS^2$ asymptotics.  
To achieve the decoupling, 5d M-theory limit of the effective 4d type IIA multicenter geometries, one wants to take the strong coupling limit of IIA string theory where an extra circle becomes geometrical; at the same time one wants to take the low-energy limit to focus on the near-source geometry.  Let the M-theory circle to be parametrized by $x_4$, with radius $R$ (this is actually the circle parametrized by the coordinate $\psi$ of the 6d solutions discussed above).
The scaling limit sends the 5d Planck scale $\ell_5\to 0$, and $R/\ell_5\to\infty$, while keeping fixed the size of the compactification $V_6/\ell_5^6$.  One also wants to keep stretched strings in the effective dynamics; in the limit these become M2 branes stretching between the charge sources and also wrapping the M-theory circle.  This sets the scaling of the brane locations to be
\be
\label{5dscaling}
y^i = \ell_5^{3} \sy^i \quad,\qquad H = \ell_5^{-3/2} \sH
\ee
where $\sy^i$ and $\sH$ are kept fixed in the limit.
This decoupling limit sets the constant terms in all the harmonic functions to zero, except for $h_0\to \frac14{R^{3/2}}$. 
This limit is entirely analogous to the M-theory limit of D0-brane matrix theory, where the excitations of the off-diagonal elements of the matrices also represent membranes wrapping the M-theory circle in an approach that starts from the dynamics of D0-brane charge centers, and the energetics of these excitations keeps them in the spectrum in the scaling limit.

Following%
~\cite{Balasubramanian:2008da}, 
the 5d metric, gauge field, and K\"ahler scalars can be written in a form very similar to~\pref{6dMetric}
\bea
\label{5dmetric}
ds^2_{5d} &=& \frac{1 }{ \sQ^{2}} \Bigl[ -(\sH^0)^2(dt+\omega)^2-2\LdB (dt+\omega)(d\psi+\omega_0) 
+ \Sigma^2(d\psi+\omega_0)^2\Bigr]
+ \sQ \, d \sy ^i d \sy ^i
\nonumber\\
A_{5d}^A &=& -\frac{\sH^0 X^A}{\sQ^{3/2}} (dt+\omega) +\frac1{\sH^0}\left(\sH^A-\frac{\LdB\, X^A}{\sQ^{3/2}}\right)(d\psi+\omega_0) +\AA^A_d
\nonumber\\
Y^A &=& \frac{2^{\frac13}X^A}{\sQ^{1/2}}
\eea
with $\psi$ parametrizing the M-theory circle, and
\bea
d\omega_0 &=& \star d\sH^0
\nonumber\\
d\AA^A_d &=& \star d\sH^A
\nonumber\\
\star d\omega &=& \langle d\sH,\sH\rangle
\nonumber\\
\Sigma^2 (\sH^0)^2 &=& \sQ^3 - \LdB ^2
\nonumber\\
\LdB  &=& \sH_0(\sH^0)^2 +\coeff13 C_{ABC}\sH^A \sH^B \sH^C-\sH^A\sH_A\sH^0
\nonumber\\
\sQ &=& (\coeff13C_{ABC} X^AX^BX^C)^{2/3}
\nonumber\\
C_{ABC}X^AX^B &=& -2\sH_C\sH^0 + C_{ABC}\sH^A\sH^B
\eea
(here $\star$ denotes Hodge star in the $\bR^3$ parametrized by $\vec y $, and $C_{ABC}$ is the triple intersection of two-cycles).

The idea is to start with charge centers that themselves have `zero entropy' and thus no internal degrees of freedom, and quantize the collective motion of these objects.  For example, the half-BPS charge $\Gamma = (1,p/2,p^2/8,p^3/48)$ is the spectral flow of a single D6-brane wrapped on $\bT^6$ and thus carries no entropy at low energies.

The quantization of the brane collective motion on the open string side is described by quiver quantum mechanics; the lightest open string degrees of freedom consist of the $U(1)$ vector multiplets describing the center of mass motion of the charge centers, together with hypermultiplets describing open strings stretching between these primitive brane bound states.  The near-horizon M-theory scaling limit will involve simultaneously taking the energy scale and brane separation to zero keeping a suitable dimensionless combination fixed.  When the branes are not coincident, the hypermultiplets are massive and can be integrated out, leading to an effective QM on the moduli space of charge centers~\cite{Denef:2002ru,Denef:2007vg}.


\subsection{Quiver QM on the Coulomb branch}

The quiver dynamics has both a Coulomb branch and a Higgs branch.  The Coulomb branch dynamics describes the motion of a set of primitive (zero-entropy) objects in the ambient $\bR^3$ parametrized by the ${\vec \sy}_a$, $a=1...N$, which are bound together by the electric and magnetic field sourced by the brane charges.  These independent motions become confined on the Higgs branch by the condensation of strings stretching between the brane centers; these states seemingly have all the `primitive' branes co-located at a single point in $\bR^3$.

Quantization of the BPS Coulomb branch spectrum has been achieved via methods of geometric quantization%
~\cite{deBoer:2008zn}.   In the quiver construction, the symplectic form for the charge center motion boils down to 
\be
\Omega = \frac14 \sum_{a\ne b} \langle \Gamma_a, \Gamma_b \rangle \frac{\epsilon_{ijk} ( \sy_{ab}^i \delta \sy_{ab}^j \delta \sy_{ab}^k) }{\sr_{ab}^3}
\ee
subject to the constraints
\be
\label{BubbleEqs2}
\sum_{a,~a\ne b} \frac{\langle \Gamma_a, \Gamma_b \rangle}{\sr_{ab}} = \langle h, \Gamma_a \rangle  
\ee
which are essentially the bubble equations; here, they come from demanding the vanishing of the effective potential that arises from integrating out the hypermultiplets.  The symplectic form is non-degenerate on the $2N-2$ dimensional solution space of the constraints and suitable for a geometric quantization approach.
The geometric quantization of the phase space using the K\"ahler form associated to this symplectic form enumerates the BPS states.
A nonrenormalization theorem supports the notion that the ground states of the quiver should match those of supergravity, and thus for the BPS states one should find the same symplectic form from the space of BPS supergravity solutions.


The two-center dynamics is rigid, in that the constraint equations~\ref{integrability} fix the center separation in terms of the charges:  
\be
\sr_{12} = \frac{\langle h,\Gamma_1\rangle}{\langle \Gamma_1,\Gamma_2\rangle}
\ee
The remaining degrees of freedom comprise the two-sphere of orientations of $\sy_{12}$, which when quantized as a phase space yields the expected $2|J|+1$ states, where $J=\half\langle\Gamma_1,\Gamma_2\rangle$.

In the three-center configuration, there are four moduli.  One of these is the magnitude $j=|\vec J|$ of the angular momentum, another is the conjugate variable $\sigma$ rotating the system around the axis of $\vec J$, and two more coordinates $(\theta,\phi)$ specify the orientation of $\vec J$; the symplectic form reduces to
\be
\Omega = -d(j\cos \theta)\wedge d\phi - dj\wedge d\sigma~.
\ee
For the centers to approach one another, $j\to 0$.  A careful analysis of the bubble constraints~\cite{deBoer:2008zn} shows that the phase space is compact and that the angular momentum lies in a range $j_- \le j\le j_+$, with scaling solutions corresponding to $j_-=0$.  K\"ahler quantization leads to a spectrum of states $\psi_{n,m}(j,\theta)$ where the quantum numbers label the number of nodes in $\sigma$ and $\phi$, 
where in the scaling case one has
\be
0\le n\le j_+-1
\quad,\qquad
-n\le m+\hf \le n  ~.
\ee
The probability density for $j$ near $j=0$ in the state $\psi_{n,m}$ turns out to vanish as $j^{2n+1}$, independent of $m$.  Thus the geometry is effectively capped, as the scaling limit is suppressed.  In the supergravity regime $j_+\to\infty$ the probability density for $j$ at fixed $n$ tends to
\be
\lim_{j_+\to\infty} |\psi_{n,m}(j)|^2 = 4 j\, e^{-2j} 
\ee
The expectation value of $j$ in this state is $\langle j\rangle = 1$.

The striking aspect of this result of~\cite{deBoer:2008zn} is that, when one considers the structure of the lowest angular momentum state, one finds that the wavefunction for the brane separation is peaked at a finite value, and vanishes as the branes are brought into coincidence.  In effect, there is a sort of angular momentum barrier which prevents the branes from lying on top of one another, and keeps the geometry effectively capped.

One might worry that the appearance of this angular momentum barrier is a consequence of the quantization of total angular momentum, and that when more centers are included there will be subsystems with $J=0$ that will be able to collapse together to form an infinite throat.  The analysis of~\cite{deBoer:2009un} shows that the individual contributions $\vec J_{ab}$ are separately quantized, not just the total, and this supports the whole collection of scaling centers against complete collapse to coincidence.

This is an entirely quantum effect -- classically, any brane configuration satisfying the constraints\pref{BubbleEqs} is allowed, including those with coincident branes.  Classically, there is a scale symmetry which sends 
\be
\sr_{ab} \to \lambda \sr_{ab}
\ee
for a cluster of centers $a,b\in\SS$,
and so one can scale the brane separations to be arbitrarily small.  Going back to the classical solution~\pref{5dmetric}, the geometry develops a throat whose redshift grows without bound as the centers approach one another.  Remarkably, quantization of the phase space shows that the states on the Coulomb branch have wavefunctions that are peaked at finite separation, and vanish in the region where an arbitrarily deep throat would develop.

Ref.~\cite{deBoer:2008zn} also estimated the size of the excitation gap in the geometry with the effectively bounded separation of the charge centers exhibited in the quiver construction, and found that it scales as $1/c$, where $c$ is the central charge of the CFT dual to the geometry.  In other words, a proper quantization of the BPS solution space leads not only to a capping off of the horizon, but also to the expected gap of the near-BPS spectrum.  This structure of the geometry can be understood at the level of linearized perturbations from the fact that the geometry with finite separation of the charge centers caps off -- the geometry ends smoothly at the bottom of the throat at a redshift value commensurate with the expected excitation gap, and the small fluctuation operator in this background has a maximum redshift of order the gap.  Smoothness of the geometry, together with a deep throat and a small excitation gap, makes this solution a promising candidate for a black hole microstate geometry.

In terms of the 5d geometry represented by the quiver quantum mechanics, and the 6d geometry dual to it, this result is quite remarkable.  Whereas classically one can have center separations going all the way down to zero, and thus an arbitrarily deep throat that can hold any amount of entropy, quantum mechanics maintains a delicate coherence of the wavefunction over macroscopic distances that keeps this horizon from forming; and provided one doesn't excite the geometrical cap too strongly, it seems that this quantum coherence will be maintained.  It seems too much to hope that this coherence will be maintained under the influence of strong local perturbations such as occur upon infall; the naive expectation would be that the infalling object decoheres these delicate correlations that are required to be maintained over macroscopic distances; a closed trapped surface forms, and the throat collapses into a singularity behind a horizon.

While this resolution of the null singularity near the horizon of the extremal geometry is welcome, it has the disturbing property that one is invoking quantum effects that are acting coherently over macroscopic distances in the geometry.  The obvious question that comes to mind is, how `real' are these coherent effects, what physical mechanism arranges them, and why are they not destroyed by interaction with local degrees of freedom?  Usually, quantum correlations over macroscopic distances are rapidly decohered through interaction with the local environment, and so one might wonder why the specially tuned BPS state is stable under even modest perturbations.%
\footnote{The quantum coherence/decoherence of macroscopic geometry is also puzzling in the context of inflation and particularly eternal inflation, where one is trying to make sense of the coherence or lack thereof of the quantum state of geometry on superhorizon scales.}


\subsection{Comments on the Higgs branch}

When the primitive branes do coincide in the transverse $\bR^3$, the stretched string hypermultiplets become massless and can condense, massing up the vector multiplets.  The resulting Higgs branch moduli space turns out to have an exponential density of states, describing an additional sector of microstates typically with parametrically larger entropy than the Coulomb branch states discussed above.  

There are actually two classes of Higgs branch states.  In the parts of the Coulomb branch wavefunction near coincident centers, the hypermultiplets of stretched strings are not so heavy, and it is not so clear that they can be integrated out.  Indeed, there is an equivalent Higgs branch representation of the Coulomb branch states where one integrates out the vector multiplets describing the center collective coordinates rather than the hypermultiplets -- the Coulomb branch wavefunctions have an echo on the Higgs branch%
~\cite{Denef:2002ru,Manschot:2010qz}, 
and so in the regime of interest these states are neither purely Higgs nor purely Coulomb, but rather can be seen from either perspective.

There are also `pure Higgs' states%
~\cite{Bena:2012hf,Lee:2012sc}, 
carrying zero angular momentum (so no barrier preventing the branes from colliding), where the hypermultiplets are fully condensed, and the vector multiplet masses are large enough that the Coulomb branch wavefunction is exponentially suppressed rather than of power law decay.

Since the branes are all coincident in the pure Higgs states, naively the geometry does not seem to be capped off and the throat seems to be infinitely deep, with a horizon.  What does this mean for the microstate geometry program?  After all, an infinitely deep, smooth throat can in principle store vastly more entropy than appears in BPS state counting, and naively the excitation gap goes to zero in contradiction to the structure of field theory duals in finite volume.

In the truncated quantum mechanical system, this issue is avoided because one has truncated the system to a finite set of degrees of freedom, and even the Higgs branch phase space that opens up at the bottom of the throat has finite volume and so there are only finitely many states, though many more than exist on the Coulomb branch.  The geometry however has many more degrees of freedom lying at the bottom of the throat, and one must find out how they are self-consistently truncated to the finite number with finite entropy that are the truly independent degrees of freedom of the black hole.  Nevertheless, if pure Higgs states are relevant to the dynamics, it would be a major blow for the microstate geometries program, since one would conclude that they vastly outnumber the Coulomb branch states, yet one would have no geometric understanding of their number or structure.


Another important difference exists between the quiver quantum mechanics construction and the 6d microstates of interest here.  The scaling limit that leads to a two-dimensional conformal field theory, dual to the $AdS_3$ near-horizon geometry of the D1-D5 duality frame, is different from the scaling limit~\pref{5dscaling} that leads to quiver quantum mechanics; one should ask whether the light degrees of freedom responsible for the structure of a given class of BPS states in one duality frame are the ones responsible in another frame.

The 4d decoupling limit of the brane dynamics involves taking $\lstr\to 0$ keeping appropriate dimensionless combinations of the torus moduli, energy, and charge center separations fixed; in other words, it is the standard Maldacena limit%
~\cite{Itzhaki:1998dd}.  
The further 5d M-theory limit involves a further scaling down of the energy and brane locations%
~\cite{deBoer:2008fk}.  
In quiver quantum mechanics, the M-theory limit fixes $E \ell_5^3/(R\Delta \sy)$ which is the energy of M-branes stretching between the charge centers; and also holds fixed $y/\ell_5^3$ and $H/\ell_5^{3/2}$ as well as $R, t,  \psi, R_5...R_{10}$, and $\Gamma_i$ in units of the characteristic energy scale $E$, while taking the 5d Planck length $\ell_5\to0$.

These two limits are similar in many respects to the decoupling limits of D0-brane quantum mechanics.  There, the standard Maldacena scaling limit for D0 branes takes 
$\lstr\to 0$ with 
$\gym^2 = \gstr \lstr^{-3}$ fixed.  The thermodynamics describes D0-brane black holes in IIA string theory on $\bR^{9,1}$.
The M-theory limit, where the typical states are approximations of black holes in M-theory on $\bR^{10,1}$, involves taking the energy scale (in units of the gauge coupling) to be of order the inverse D0 charge $N$ in the large $N$ limit, and it is only in this further limit that the M-theory circle becomes effectively large.  One now scales the D-particle spacing and energies relative to 11d Planck units, where $\lstr^2=\lpl^3/R$ and $R$ is the radius of the M-theory circle.


The scaling limit just described differs from the scaling limit that leads to 6d D1-D5-P microstate geometries.  As discussed in section~\ref{sec:Quivers}, this limit starts with the 5d theory with three sets of intersecting M2-branes in M-theory on $\bT^6$; after shrinking the two-cycle (say in directions 9-10) wrapped by one set of M2-branes to well below the 11d Planck scale, the appropriate type IIB duality frame has D3 branes intersecting over the type IIB circle dual to the shrunken torus, and the M2 branes wrapping the shrunken torus dualize into momentum along the IIB circle whose radius is 
$\tilde R_9 = \lpl^3/(R_9R_{10})$.  
In this duality frame, the decoupling limit that leads to $AdS_3\times\bS^3$ fixes
\be
\label{6dscaling}
\tilde R_9 \sim \lstr^0 \quad,\qquad  {R_5 R_6 R_7 R_8}\sim {\lstr^4} 
\quad,\qquad y^i \sim \lstr^2  ~,
\ee
where again dimensionful quantities are referred to the characteristic energy scale of the system.
This limit differs from~\pref{5dscaling}, where the energy cost of each set of the triplet of M2-brane charges scales the same way.
Instead,~\pref{6dscaling} keeps the momentum along $\tilde R_9$ as a light excitation in the effective theory, naively as light or lighter than the stretched strings (hypermultiplets) of the quiver quantum mechanics, rather than treating it as part of the heavy background charges.  Because the scaling limits are different, features of the geometry that are not resolved by the degrees of freedom kept in quiver quantum mechanics might instead be resolved by the behavior of these new light excitations of the 6d theory, which in the 5d scaling limit are frozen as part of the heavy background.

In the typical 6d microstate, the geometry is varying along both the $v$ and $\psi$  directions; however, as was mentioned above, only when there is an isometry along the circle being dualized is there a simple, direct relation between the harmonic functions of the geometry in different duality frames.  This excludes the vast majority of microstate geometries; they will not be described by quiver quantum mechanics.  The  $v$-dependence of the generic three-charge background breaks this symmetry, and complicates the relation between the BPS spectra of the quiver QM and the 6d geometry, and in particular the issue of whether the geometry is capped off at finite radius.  

The authors of~\cite{Bena:2014qxa} have suggested that because 1/3 of the central charge of the CFT dual comes from (fermionic) degrees of freedom carrying angular momentum on the $\bS^3$ as well as momentum along the $v$ circle, there will be a distribution of momentum and angular momentum along the microstate geometry, and this ensemble of $(v,\psi)$ dependent states will have their centers supported against collapse as in the three-charge example.  It is hoped that in this way, an order one fraction of the entropy will be accessible as distinct geometries. This scenario assumes that there isn't a mechanism that engineers charge/spin separation as is known to occur in certain condensed matter systems%
~\cite{Luttinger:1963zz,Haldane:1981zza}; 
such a mechanism might allow the angular momentum to be carried by a `halo' while most of the entropy is carried by other degrees of freedom on the inner horizon (see below).  One might worry that if there is a vastly larger entropy in the Higgs branch, that the system may try to perform such a separation.  This then leads to a puzzle about how these degrees of freedom are to recombine to make Hawking radiation if they are so distantly separated.

The observation that there are light degrees of freedom in 6d black holes (the $v$ and $\psi$ dependence of the geometry) that are not accounted for in the quiver quantum mechanics, does not necessarily mean that the hypermultiplets that generate the Higgs branch in 5d are irrelevant in 6d.  One should in particular understand what becomes of the exponential density of pure Higgs states, which in explicitly checked examples vastly exceeds that of the Coulomb branch states.

The picture of the Higgs branch gleaned from the quiver {quantum mechanics} seems at odds with the understanding of generic black hole states gleaned from the ensemble geometry, which from the discussion of section~\ref{sec:Thermo} indicates that excitations above extremality extend out substantially into the inter-horizon region.  Instead, in the pure Higgs states, the vector multiplet wavefunction dies off exponentially rapidly away from $r=0$, which naively should be the horizon -- the {\it inner horizon}, if the thermodynamics and the covariant entropy bound are to be believed.  But then one is concentrating the bulk of the black hole degrees of freedom in a region causally separated from the black hole exterior by a macroscopic amount, which only grows as the black hole is further excited.  There would have to be an additional form of nonlocality in the theory in order to avoid the usual information paradox trap when trying to extract information from the black hole through Hawking radiation.%
\footnote{Such nonlocalities in the effective theory have been advocated for instance in~\cite{Giddings:2012gc}.}

To summarize, the Higgs branch of quiver quantum mechanics has a vast reservoir of states, larger than the spectrum of Coulomb branch states.  These states carry no angular momentum, and their wavefunction is supported at $r=0$ where naively the throat is infinitely deep, and so it looks like a horizon has formed.  If this result carries over to the 6d type IIB geometries obtained after dualization from M-theory in 5d, then necessarily the bulk of the microstates are not realized as capped geometries.
If the Higgs branch states are indicative of the structure of the majority of the 6d BPS spectrum, the considerations of section~\ref{sec:Thermo} argue that these states should be associated with the inner horizon; however their wavefunction seems not extend into the inter-horizon region, if the quiver QM wavefunctions are an accurate guide.  Of course, one should also remember that the form of wavefunctions is not protected by any nonrenormalization property, so the wavefunctions in the quantum mechanics may be a poor guide to the structure of the 6d theory.

The Coulomb/Higgs terminological distinctions we have been making are probably an expedient (and perhaps misleading) fiction that glosses over a more subtle truth.  In quiver quantum mechanics, the Coulomb branch states can have an echo on the Higgs branch and vice versa.  There is reason to suspect that the distinction is even more subtle in any formulation relevant to 6d geometries.  Further insight into the nexus between the two, and how communication takes place across it, would certainly be welcome.

If the pure Higgs states of the quantum mechanics are somehow irrelevant, part of the justification ought to come from understanding the analogue of the hypermultiplets of the quiver in the 6d geometry.  They start off life as strings stretching between primitive D-brane bound states involving D6-branes in 4d string theory.  In the M-theory limit of the quantum mechanics, these strings become M2-branes wrapped on the M-theory circle and stretching between KK monopoles; in other words, the geometry has nontrivial two-cycles which consist of the M-theory circle fibered over the interval between centers in the Gibbons-Hawking geometry.  Under the duality to IIB, these M2-branes become D3 branes wrapping this $\bS^2$ as well as the type IIB circle dual to the shrunken $\bT^2$ in M-theory.  When one brings charge centers together in the Gibbons-Hawking base, the $\bS^2$ vanishes and the D3-brane becomes a tensionless string.  The condensation of this string then ought to be related to entering the Higgs branch of the 6d theory (or rather, the generic microstate would involve a condensate of such strings).  

More precisely, when a cycle vanishes in the Gibbons-Hawking base manifold of the 6d geometry, the various warp factors in the metric cancel that shrinkage and ensure that the cycle remains of fixed proper size (since the throat geometry approaches $AdS_3\times \bS^2$).  
Nevertheless, objects at the bottom of the throat cost little energy, because the same warp factors govern the redshift in the metric~\pref{6dMetric}, so the effect is the same as if the cycle was vanishing.  The effective string from the three-brane wrapping the vanishing cycle is not necessarily tensionless when the cycle collapses -- there are additional antisymmetric tensor field moduli of the NS B-field and RR two-form that are associated to the two-cycles.  Only when the flux of these potentials through the two-cycle vanishes does the string become truly tensionless~\cite{Aspinwall:1994ev}.
Naively it seems that this modulus is unconstrained and will be dynamical on a compact geometry, and thus the wavefunction would have support on the tensionless string limit.  We thus see no reason that the hypermultiplet dynamics of the Higgs branch will be suppressed in the 6d theory.%


\section{Fivebrane singularities}
\label{sec:Fivebranes}

It turns out that many of the potential geometrical pathologies (orbifold singularities, scaling limits, \etc) in the microstate geometries are due to the configuration of the underlying background sources, whose behavior closely parallels that of fivebranes in well-studied situations.  It will therefore be useful for us to review several facts about fivebrane dynamics, beginning with the duality between fivebranes and orbifolds (the discussion here follows%
~\cite{Harvey:2001wm}, section 4.2).  The structure of the Coulomb and Higgs branches of fivebranes will illuminate the issues raised above, and provide further support for the notion that the capped microstate geometries do not account for the bulk of the entropy of three charge black holes.

\subsection{Fivebrane/orbifold duality}

\begin{figure}
\centerline{\includegraphics[width=4in]{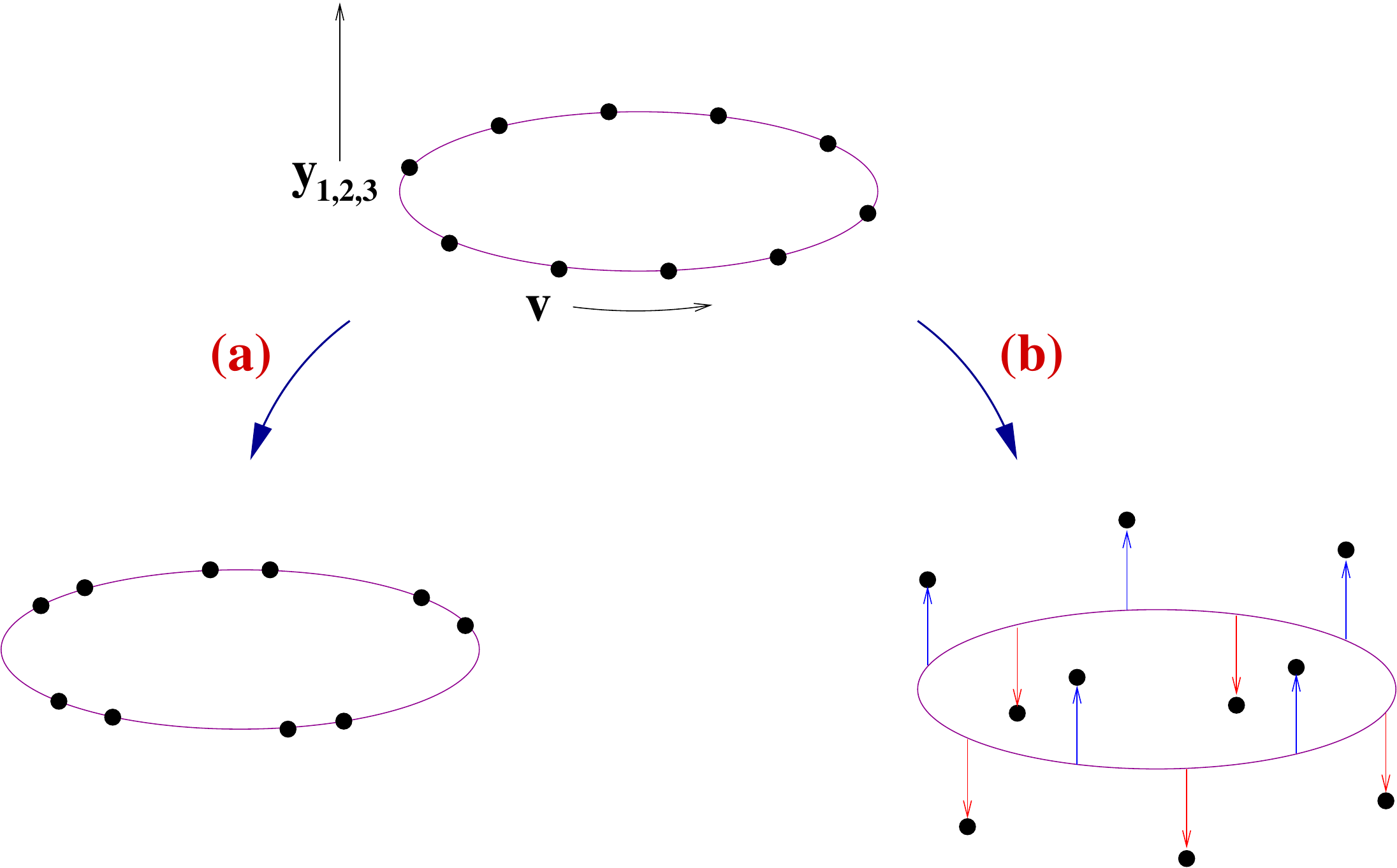}}
\setlength{\unitlength}{0.1\columnwidth}
\caption{\it 
Two perturbations of a $\bZ_{n}$ symmetric arrangement  
of type IIB fivebranes on a circle, dual to type IIA  
string theory on $\bC^2/\bZ_n$:  
(a) moving the fivebranes on $\bS^1$ is related to   
changing NS $B$-field fluxes through vanishing cycles  
on the IIA side;  
(b) moving them in $\bR^3$ is dual to turning on the triplets of  
geometrical blow up modes of the vanishing cycles   
on the IIA side.  
}
\label{fig:5braneOrbDuality}
\end{figure}

The orbifold theory  
$\bC^2/\bZ_{n}$ is  
T-dual to the theory of fivebranes on a circle,  
in an appropriate limit~\cite{Ooguri:1995wj,Kutasov:1995te}.  
Consider type II string theory on $\bR^{8,1}\times \bS^1$,  
with $n$ NS5-branes symmetrically arranged on the circle,  
which we take to have circumference $R$, and parametrized by  
$v$; and let $y^{1,2,3}$ parametrize the $\bR^3$ transverse  
to the fivebranes (see figure~\ref{fig:5braneOrbDuality}). Then in the limit  
\be
\label{decoup} 
  \gstr\rightarrow 0\quad,\qquad  
  R/\lstr\rightarrow 0\quad,\qquad  
	{\rm with} \quad \frac{R}{ \lstr\gstr } \quad {\rm fixed}  ~,
\ee  
type IIB string theory in the fivebrane background is  
equivalent to type IIA string theory on the orbifold  
$\bC^2/\bZ_{n}$ (and vice versa). The two descriptions are related by T-duality applied to  
the circle parametrized by $v$. 

The orbifold has $n-1$ hypermultiplets of moduli  
coming from twisted sectors;  
the four real parameters in each hypermultiplet consist of   
the NS $B$-field flux through one of the $n-1$ vanishing  
cycles of the orbifold ALE space, together with a triplet  
of modes that blow up that cycle.  
The $B$-flux is a periodic coordinate, while the   
blow up modes parametrize $\bR^3$.  
These map on the fivebrane side into the relative locations  
of the fivebranes on the $\bS^1$ and $\bR^3$, respectively.  
The standard $\bC^2/\bZ_{n}$ orbifold CFT 
corresponds to the point in moduli space where the  
fivebranes are coincident in $\bR^3$ and symmetrically  
arranged on the $\bS^1$ (as in the top of figure~\ref{fig:5braneOrbDuality}).   
The $\bZ_n$ symmetry that cyclically permutes the fivebranes  
is the $\bZ_n$ quantum symmetry of the orbifold CFT.
  
Near-coincident NS5-branes generate a target space for  
perturbative worldsheet string theory which develops  
a throat along which the string coupling grows;  
the throat becomes infinitely long, and the coupling at its end  
diverges, in the limit where fivebranes coincide~\cite{Callan:1991dj}.  On the IIA side, this singularity of  
the worldsheet CFT can be understood from considerations  
of linear sigma models, in the limit where the worldsheet theta angle is turned off.  
 
One can also match the structure of D-branes on the two sides.  
The limit~\pref{decoup} keeps fixed the mass in string units  
of D1-branes stretching between the NS5-branes on  
the IIB side; their mass scales as   
\be
\label{mdone}  
 \lstr m_{\sst W}=\frac{R}{n\;\lstr \gstr^{\sst B}}  
\ee  
at the point in moduli space related to the orbifold.  
D1-branes of fractional winding are pinned to the NS5-branes  
they begin and end on, while D1-branes of integer winding  
are free to move in the $\bR^3$ transverse to the NS5-branes.  
Exactly the same structure is obtained in IIA string  
theory on $\bC^2/\bZ_{n}$. There, fractional  
D0-branes of the orbifold are the W-bosons of a spontaneously  
broken 5+1 dimensional gauge symmetry localized on the orbifold  
singularity; their mass is 
\be
\label{mfrac}  
  \lstr m_{\sst W}=\frac1{n\;\gstr^{\sst A}}
  \quad,\qquad
  \gstr^{\sst A}=\gstr^{\sst B}\lstr/R\ .
\ee
These excitations are D2-branes wrapping  
the vanishing cycles of the ALE space, and carrying  
a fractional unit $1/n$ of D0-brane charge.  
  
 
  
Fractionally wound branes become massless if fivebranes coincide  
(IIB), or equivalently (IIA) when the B-flux through vanishing cycles  
of the ALE space is turned off~\cite{Aspinwall:1994ev};   
the D-brane gauge dynamics then becomes strongly coupled.  
This is the open string reflection of the singularity  
of the closed string sector noted above. 

A similar structure arises for NS5-branes on $\bR^4$ rather than $\bR^3\times\bS^1$.  On $\bR^4$ one has the CHS construction~\cite{Callan:1991dj}, which has been studied in great detail from a more modern perspective in%
~\cite{%
Aharony:1998ub,%
Giveon:1999zm,%
Giveon:1999px,%
Giveon:1999tq%
}.  Fivebranes separated on the Coulomb branch make a throat that is smoothly capped off as seen by short (fundamental) strings.  A long throat with large redshift develops as the fivebranes approach one another along the Coulomb branch; the depth of the throat is controlled by the brane separation.  New light (and strongly coupled) degrees of freedom -- again D-branes stretching between the fivebranes -- arise in the limit that the branes collide~\cite{Elitzur:2000pq}.  The depth of the throat is directly tied to the lightness of these degrees of freedom, which are associated to the `little strings' of fractionated tension that inhabit coincident fivebranes.  

We claim that similar phenomena occur in the present context, and that one can understand the appearance of a large redshift when a black hole is forming in $AdS_3\times\bS^3$ as arising from background sources that are approaching one another, revealing new light brane excitations.  Analogues of both of the above situations involving fivebranes arise in the context of three charge systems.  First, if the Gibbons-Hawking base $\BB$ has charge centers with greater than unit charge, $|q_a|>1$, the base has an orbifold singularity whose dynamics parallels that of fivebranes on $\bR^3\times\bS^1$.  Scaling solutions, where charge centers can approach one another arbitrarily closely, are the analogues of fivebranes on~$\bR^4$.  The fact that the entropy of fivebranes is accounted for by the Hagedorn entropy of `little strings' on the Higgs branch rather than by quantizing excitations of the cap on the Coulomb branch, suggests that a similar fate should await the three charge capped microstate geometries of the three-charge system.


\subsection{Singularities in D1-D5 microstate geometries}

The above structure already appears in the two-charge backgrounds of the D1-D5 system.  
The chiral primaries of this theory can be mapped to an F1-P duality frame where the charges are simply winding and momentum of a fundamental string%
~\cite{Lunin:2001fv,Lunin:2002bj,Martinec:2002xq} (see~\cite{Mathur:2005zp} for a review).  
After smearing the source over the (dual of the) $v$ circle and dualizing back, explicit expressions for the supergravity fields are obtained for an arbitrary quantized profile $X^i(v)$ of the string oscillation in the base $\BB=\bR^4$ (\ie\ a single pole with unit residue in the Gibbons-Hawking parametrization of $\BB$).  One finds the coefficient functions in the metric~\pref{6dMetric}
\bea
\label{LMGeoms}
Z_1 &=& 1 + \frac{Q}{L} \int_0^L \frac{dv}{(\xx - \X(v))^2}
\quad,\qquad
Z_2 = 1+ \frac{Q}{L} \int_0^L \frac{(\dot \X(v))^2\,dv}{(\xx - \X(v))^2}
\nonumber\\
\nonumber\\
\beta &=& (A+B)/\sqrt{2}
\quad,\qquad
k = (A-B)/\sqrt{2}
\quad,\qquad
\FF = 0
\\
\nonumber\\
A_i &=& -\frac{Q}{L}\int_0^L \frac{\dot X^i(v) \,dv}{(\xx - \X(v))^2}
\quad,\qquad
dB = *dA
\nonumber
\eea

\begin{figure}
\centerline{\includegraphics[width=3in]{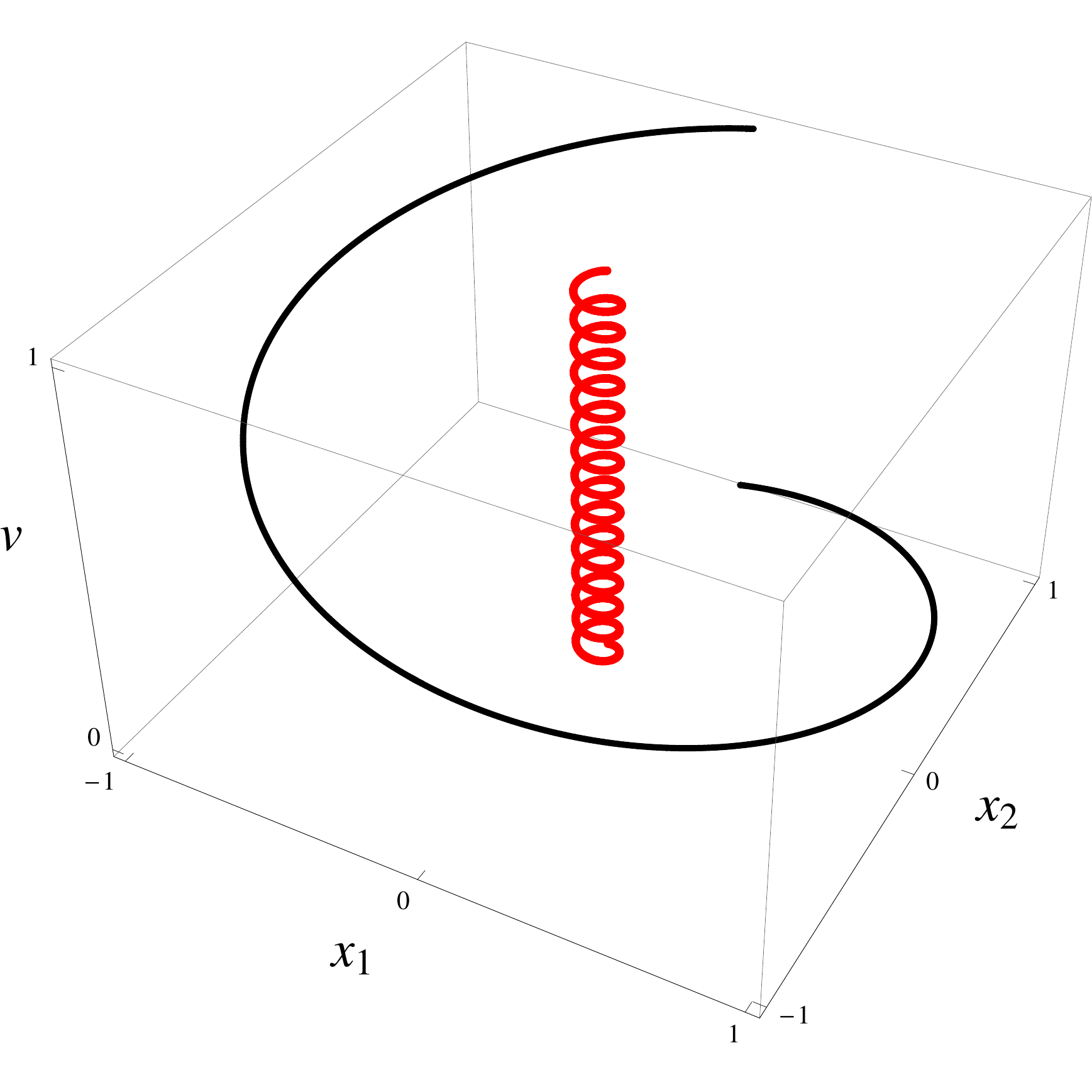}}
\setlength{\unitlength}{0.1\columnwidth}
\caption{\it 
Sources in equation~\pref {FPSource}  for the two-charge solution.  Putting a macroscopic number of quanta in the lowest mode (the laconic source shown in black, making a single turn in the $X_1$-$X_2$ plane as one moves along the $v$ circle) constitutes a macroscopic ring source whose geometry turns out to be the spectral flow of the global $AdS_3\times\bS^3$ geometry.  Putting a single quantum in the highest mode (the tightly coiled spiral shown in red) makes an orbifold geometry
$(AdS_3\times \bS^3)/\bZ_{n_1n_5}$.
All the two-charge BPS geometries are specified by such a coiling long string source, which when separated in space describes a state on the Coulomb branch of D1-D5 system.}
\label{fig:TwoExtremes}
\end{figure}

Perhaps the simplest choice for $\X(v)$ is to take
\be
\label{FPSource}
X_1 + i X_2 = a \, e^{i\omega v}
\quad,\qquad
X_3+i X_4 = 0
\ee
for the four noncompact coordinates of the base $\BB=\bR^4$ transverse to the $v$ circle (and the $\bT^4$), with the string wound $n_5$ times over the $v$ circle of radius $R$, and carrying all its momentum excitations in the $k^{\rm th}$ oscillator mode.  Translated to the D1-D5 frame, one has
\be
\omega = \frac{k R}{n_5}
\quad,\qquad
a =\frac{\sqrt{Q_1Q_5}}{kR} 
\ee
as the image of the parameters characterizing the state.  Such a string source is depicted in figure%
~\ref{fig:TwoExtremes}.

As shown in~\cite{Lunin:2001fv}, the source~\pref{FPSource} with mode number $k$ generates a D1-D5 geometry $(AdS_3\times \bS^3)/\bZ_k$.  The lowest mode $k=1$ describes global $AdS_3\times\bS^3$, or more precisely the maximally spinning state obtained from this vacuum geometry by two units of spectral flow in the spacetime CFT%
~\cite{Maldacena:2000dr}.
The deepening throat with increasing $k$ is reflected in the dual source by the decrease of the ring radius $a$ by a factor of $k$, so that the strands of the string are drawn closer together in the $X_1$-$X_2$ plane.  The strands are furthermore packed more densely along the $v$ circle by a factor of $k$, as the source makes $k$ windings before returning to itself as it travels from $v$ to $v+2\pi n_5 R$.  The difference from the fivebrane story above is that now one is dealing with the underlying long effective string carrying both one-brane and five-brane charges, rather than just the five-branes; also, the two-charge BPS states generically have no moduli because the source configuration is fixed by the choice of mode excitations of the dual F1-P state.  The choice $k=n_1n_5$ makes the orbifold defect angle large, and the throat is deep, with excitation gap of order $(n_1n_5)^{-1}$; this state is very near but just below the threshold for the extremal BTZ black hole.  The long string source, which is smeared over the $v$ circle in order to perform the duality transformation, is a tiny helix whose strands are coincident in the directions transverse to $v$ but secretly just slightly separated in $v$, see figure~\ref{fig:TwoExtremes}.  Thus, just like the fivebrane situation reviewed above, deep throats are tied to underlying sources approaching one another along the Coulomb branch.  The main difference with the situation described in the previous subsection is that the source generating the deep throat whose singular limit is associated to fivebrane sources, is here replaced by the $AdS$ throat associated to long string sources.

\begin{figure}
\centering
  \begin{subfigure}[b]{0.4\textwidth}
    \includegraphics[width=\textwidth]{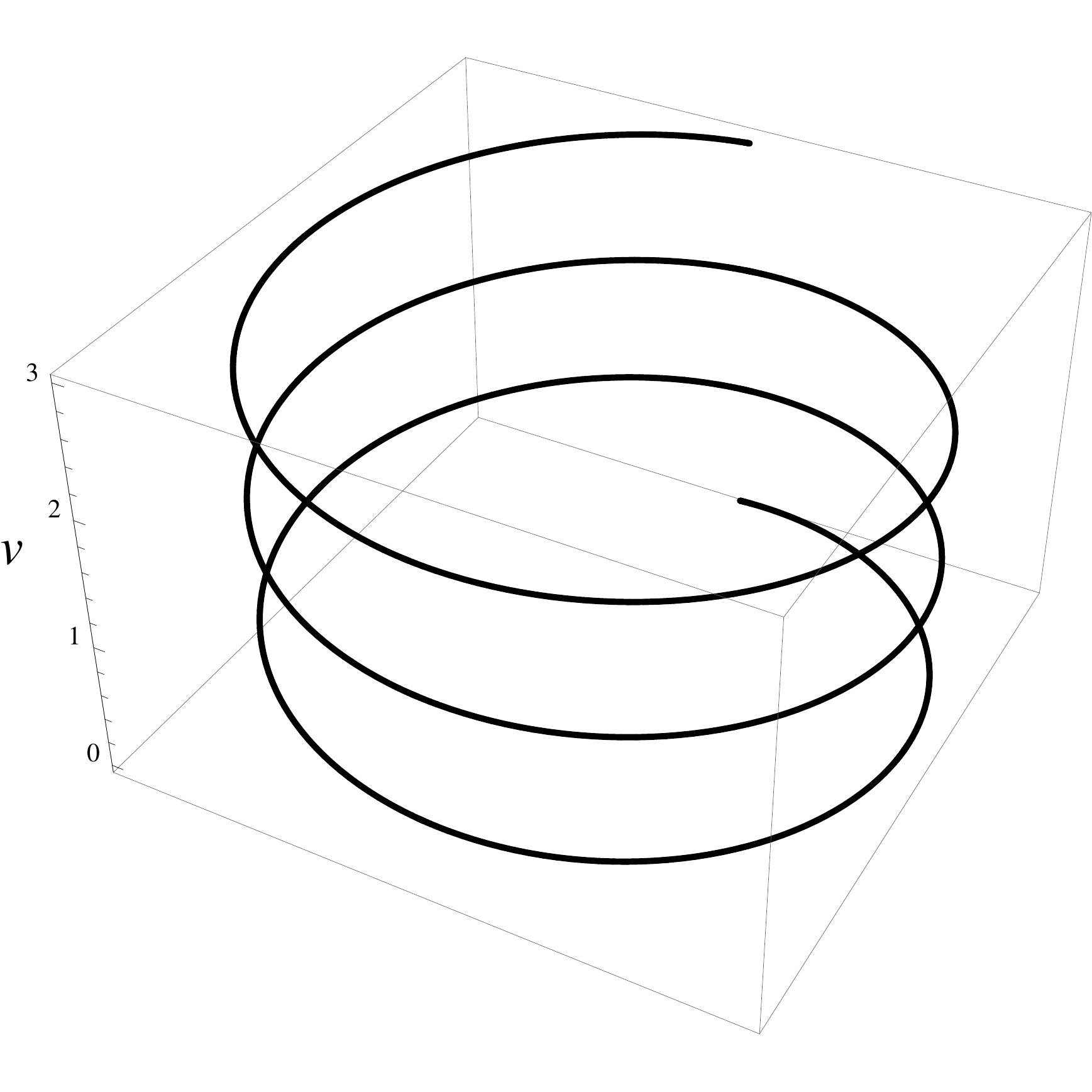}
    \caption{ }
    \label{fig:AdSchiralprimary3}
  \end{subfigure}
\qquad
  \begin{subfigure}[b]{0.4\textwidth}
    \includegraphics[width=\textwidth]{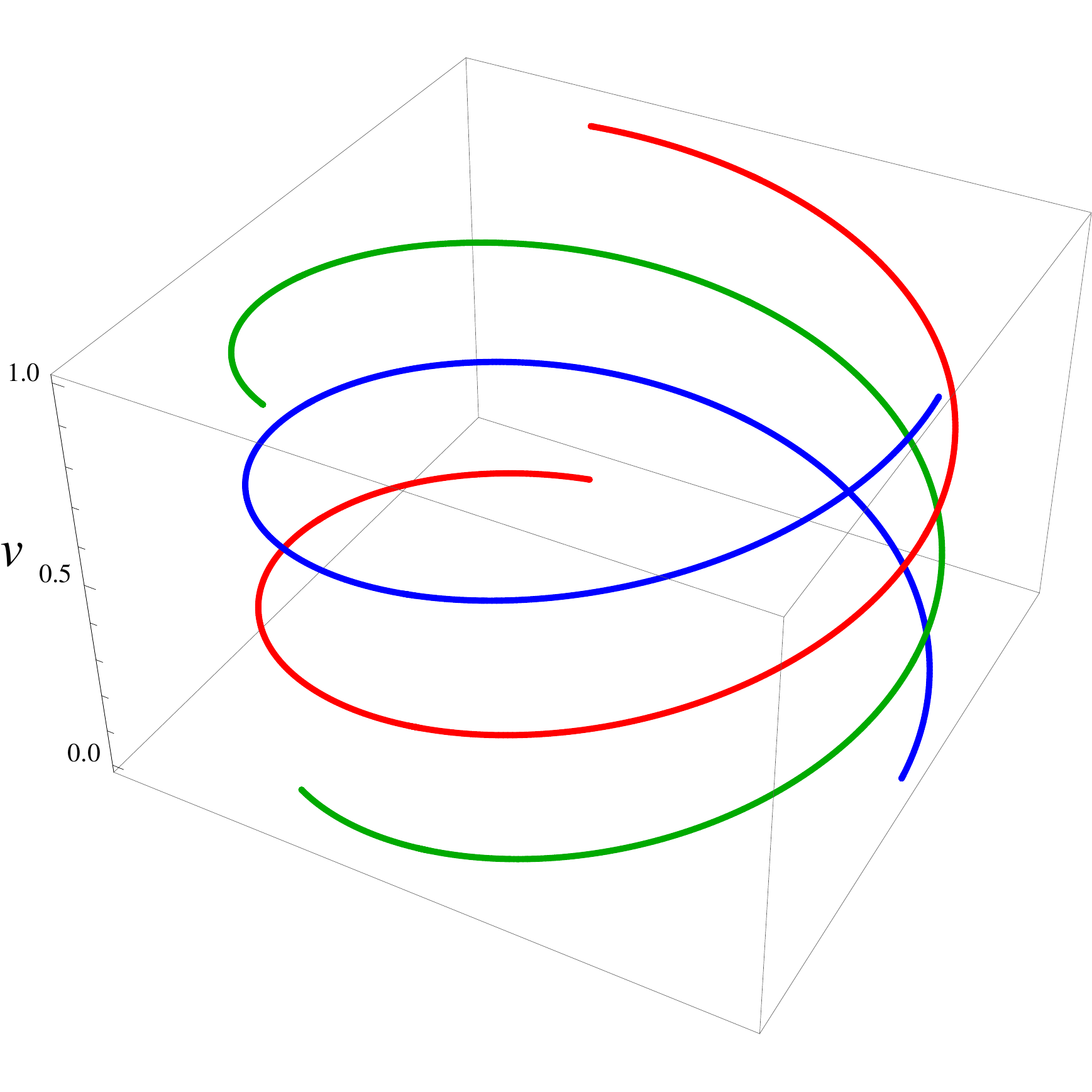}
    \caption{ }
    \label{fig:AdSorbifold}
  \end{subfigure}
%
\caption{\it 
Sources for the two-charge solution.  (a) A single BPS source with $n_5=3$ and $k=3$ has the strands of the source locked at finite separation on the covering space of the $v$ circle, but coinciding in spacetime; (b) Splitting the single string into three string sources with $n_5=1$ and $k=1$, and separating them, desingularizes the coincident source singularity that arises when $k$ and $n_5$ have a common divisor in the single-string source. 
}
\label{fig:AdSorbs}
\end{figure}

A closely related $(AdS_3\times \bS^3)/\bZ_m$ orbifold is described in~\cite{Martinec:2002xq}.  In the language of%
~\cite{Lunin:2001fv}, when $k$ and $n_5$ have a common divisor the configuration is singular, because the source traces over the same curve in spacetime $m={\it gcd}(n_5,k)$ times.  The example $n_5=k=3$ is shown in figure%
~\ref{fig:AdSchiralprimary3}.  One can desingularize the geometry by splitting the source into $m$ separate string sources, each carrying mode number $k/m$, and separating them along the $v$ circle, as shown in figure%
~\ref{fig:AdSorbifold}.  Placing the $m$ strings in a $\bZ_m$ symmetric arrangement leads to the background worked out in~\cite{Martinec:2002xq}, which showed explicitly how the moduli of the orbifold $(AdS_3\times \bS^3)/\bZ_m$ map to the locations of the sources in the \hk base of the geometry~\pref{6dmetric} in the construction of%
~\cite{Lunin:2001fv,Lunin:2002bj}.%
\footnote{As shown in~\cite{Seiberg:1999xz}, the NS duality frame with all Ramond moduli turned off is a singular point in the moduli space of the spacetime CFT, where brane charge can escape to the boundary of $AdS_3$.  This singularity is regularized by turning on these moduli, which generate an attractive potential between the branes which lifts the flat directions of the orbifold.  The worldsheet description used in~\cite{Martinec:2002xq} is at the singular point of the moduli space, where the brane separations are a true flat direction of the configuration space.}
Thus the construction of~\cite{Martinec:2002xq} realizes a variant of the fivebrane-orbifold duality depicted in figure~\ref{fig:5braneOrbDuality}.

Thus we see that when orbifold singularities arise in the \hk base $\BB$ of the microstate geometry, one needs to look further to see whether the geometry is actually nonsingular, or whether instead one has landed on a singular point in the moduli space.  The quantity that governs the distance to the singularity in these two-charge background configurations is the source separation, which governs how close one is to a singularity of the effective theory. In the special symmetric configurations above, multiple strands of the source travel the same path in the noncompact $\bR^4$ , parametrized by $\psi$ (the angular direction in the $X_1$-$X_2$ plane).  The sources are only separated as they wind along a cycle on the $\bT^2$ parametrized by $v$ and $\psi$, and if the source wraps that cycle multiple times, there is a singularity.  The transverse separation of the strands governs how close one is to the singularity, \ie\ the tension of the `W-branes' that stretch between the sources. The picture above indicates that when $n_5$ and the mode number $k$ do not have a common divisor, the theory can be regular -- that the sources are separated along $v$ and the would-be angle modulus is lifted (\ie\ is a fixed scalar).

When the orbifold is not singular, it is as usual because there is nonzero NS $B$-field flux through the two-cycles of the geometry, which are collapsed at the orbifold locus.  This desingularization is hidden in the geometry but well-understood from the string theory perspective~\cite{Aspinwall:1994ev}.  In the simple solutions~\pref{FPSource}, the configuration of the string source tells us that this flux is nonzero, because the strands of the source are separated along the helix, at least when there are no retracings of the path.  The orbifold fixed point that comes closest to being a black hole has order $n_1n_5$, and has $n_1n_5$ species of light wrapped branes; for instance, D3-branes wrapping the vanishing cycles will be strings whose tension is of order $(\gstr\lstr^2 n_1n_5)^{-1}$, and will only become lighter if the strands of the long string are pushed closer together and the B-field is turned off.  For the generic two-charge configuration arising from dualization of an F1-P source, one expects that there will be light brane excitations in the D1-D5 geometry whenever the string source comes close to self-intersecting.  These `W-branes' signal the emergence of the long string phase.

The generic source profile $X^i(v)$ in~\pref{LMGeoms} consists of the string executing a random walk in the base $\BB=\bR^4$ as it winds along the $v$ circle, with an average mode number $\bar k \sim\sqrt{n_1n_5}$, a radius of gyration of order $\sqrt{n_1n_5}$, and a typical spacing within $\BB$ to the nearest other point on the string of order $(n_1n_5)^{1/6}$ in units of the 5d Planck length~\cite{Mathur:2005zp,Mathur:2007sc}; the fine-tuning that might cause the source to trace over the same path will be absent, and the source string is generically far from self-intersecting.   An intriguing analysis%
~\cite{Lunin:2002qf,Mathur:2005zp,Mathur:2007sc}
shows that the number of solutions that fit within the typical radius of gyration satisfies the Bekenstein-Hawking area law $S\sim A/4G$.  However these states are somewhat far from being black holes.  The typical cycle in the symmetric product  orbifold has length equal to the typical mode number $\bar k \sim\sqrt{n_1n_5}$, and so the gap in the spectrum is much larger than one expects of a solution with a truly deep throat.
The typical such state will carry a characteristic angular momentum $J\sim \sqrt{n_1n_5}$ since  the angular momentum is proportional to the number of modes; the entropy formula~\pref{Sbmpv} then says that one needs $n_p\sim 1$ in order to rise up to the BMPV black hole threshold, however to achieve this one must excite the available cycles a macroscopic amount using of order $\sqrt{n_1n_5}$ excitations down at the bottom of the throat; but even though the state will then have the same quantum numbers as a BMPV black hole, it will not have the same excitation gap and so the microstate is not a generic black hole microstate.
Instead, the two-charge geometries exhibit the long string as an explicitly visible bare source tracing out a path in $\BB$ as a function of $v$, which is then smeared over $v$.  These BPS geometries carry angular momentum, whose centrifugal force pries apart the long string, allowing us to see it as a Coulomb branch state.

A similar story to the $AdS$ orbifolds above plays out in the two-center solutions of section~\ref{sec:MicroGeom}.  In the two-center solutions~\pref{TwoPoleV}, the residues $s$ and $s+1$ of the poles in the harmonic function $V$ of the Gibbons-Hawking base are such that $s(s+1)$ is a multiple of the KK dipole charge $k_3$.  There are orbifold singularities (of order ${\it gcd}(s,k_3)$ and ${\it gcd}(s+1,k_3)$) at the north and south poles of the $\bS^2$ consisting of the $\psi$ circle fibered over the line joining the two centers, where the fiber degenerates.  These orbifold singularities are the locus of $m-1$ additional cycles which have been blown down, where $m$ is the order of the orbifold quotient.  These orbifold singularities will be benign if there is antisymmetric tensor flux through the collapsed cycles.  It would be interesting to work out the values of $B$ in this situation, which should be frozen at some particular nonzero values.


In the two-center solutions, the map to chiral primaries of the symmetric product given in~\cite{Giusto:2012yz} indicates that the excitation gap is of order $k_3^{-1}$.  If one wants the excitation gap to approximate that of black holes, one wants $k_3\sim n_1n_5$.  We then conclude that there are orbifold singularities of order $\sqrt{n_1n_5}$ or worse at the poles of the nontrivial sphere in $\BB$, when the depth of the throat is deep enough for the geometry to look like a black hole.

Solutions with three or more centers admit scaling solutions for the microstate geometries, where a cluster of poles in the Gibbons-Hawking base coalesce.  These microstate geometries represent a situation analogous to fivebranes on the Coulomb branch in $\bR^4$, since the poles are the locus of sources of the background charges.  The centers on the \hk base $\BB$ are free to move around, modulo the constraints imposed by the bubble equations~\pref{BubbleEqs}. 
Scaling a cluster of centers toward coincidence in $\BB$ is the direct analogue of moving fivebranes close together; a deep throat develops, and wrapped brane excitations that are `W-branes' stretching between charge centers, become lighter and lighter in the process.

The `spacetime foam' limit of many centers was studied in~\cite{Bena:2006is}.  Setting for simplicity $q_a=(-1)^{a+1}$ for $N=2M+1$ centers, and the dipole charges of each type all of the same order as the mean value
\be
k_a^I = \bar k^I (1+\OO(1)) ~,
\ee
one finds that in the large $N$ limit the conserved charges scale as
\be
Q_1 \sim 4N^2\bark_2\bark_3
~~,~~~
Q_2 \sim 4N^2\bark_3\bark_1
~~,~~~
Q_3 \sim 4N^2\bark_1\bark_2
~~,~~~
J_R \sim 8 N^3 \bark_1\bark_2\bark_3  ~,
\ee
with 
\be
\frac{J_R^2}{Q_1Q_2Q_3} - 1 \sim \OO(N^{-2}) ~.
\ee
The value of $J_L$ depends on the solution of the bubble equations, but was checked numerically for several examples and found to be subleading in the large $N$ limit.  Thus once again the solutions seem to be near but just below the BMPV black hole threshold.  With $N$ centers there are $N^2$ separate two-spheres, each holding of a few units of each type of charge.  By moving any given group of centers together in a scaling solution, a long throat develops and one pushes the associated charge cluster towards the Higgs branch.

In the two-charge BPS geometries, and (assuming they are nonsingular) the two-center solutions discussed above, the regularity of the solution comports with the fact that the moduli are all frozen, and there is a gap to exciting the long string degrees of freedom. More general multicenter solutions have a combination of orbifold singularities, centers that are not free to approach one another due to the bubble equation constraints, and scaling clusters.
The features of these geometries contain the information about the underlying long string that sources the geometry, which becomes the long string of the black hole spectrum as the excitation gap approaches the value typical of the black hole states.  We thus have a concrete picture of where the long string lurks in the geometrical side of the duality.  In the geometries with the deepest throats, the excitations bound to the long string do not cost a lot of energy, and small non-extremality may cause strands of the source string to approach one another, leading to a singularity in the effective field theory.  The depth of the throat, or the size of the cycles, is directly tied to how near one is to liberating some portion of the long string degrees of freedom.  It is important to realize that the effective field theory becomes singular not because the underlying theory is pathological, rather it is simply that new light degrees of freedom arise and so it was a mistake to integrate them out; our approximation scheme is what is breaking down.  Just as the singularities of fivebranes on the Coulomb branch signal the appearance of the Higgs branch of `little strings' which accounts for the black fivebrane entropy, similarly in the D1-D5 system new light degrees of freedom arise, associated to the long string (and the Higgs branch in the Coulomb/Higgs dichotomy).  In the case of fivebranes, one doesn't count the entropy of black fivebranes by quantizing the excitations in the cap of the Coulomb branch geometry.  Similarly, it is the long string, whose excitations are liberated on the Higgs branch, that we expect to be responsible for the three charge black hole entropy, rather than a consideration of distinct ways of wiggling the microstate geometry. 

In the next section, we propose that the long string degrees of freedom of the Higgs branch not only count the entropy; they also resolve the null singularity at the inner horizon of BTZ black hole geometries, and not just at extremality.  This sort of mechanism has always been the way that string theory resolves timelike singularities, via the appearance of either light perturbative string states%
~\cite{Dixon:1985jw,Aspinwall:1993nu,Aspinwall:1994ev}
or light D-branes~\cite{Strominger:1995cz};
the analysis here points to a mechanism whereby string theory also resolves null singularities in a very similar fashion.  Can spacelike singularities be far behind?  After all, the same long string structure will be operating behind the outer horizon, arbitrarily far from extremality.  In the following, we will provide a picture of how that resolution takes place as a consequence of the string/black hole correspondence principle of~\cite{Horowitz:1996nw}.


\section{Discussion and Speculations}
\label{sec:Discussion}

\subsection{What can we learn about black holes from the Coulomb branch?}

Before delving into the issue of singularity resolution, let us address the question of what can be gleaned from the microstate geometries program if it indeed falls short of accounting for three-charge black hole entropy.  We suspect that these geometries still have an important role to play in sorting out black hole structure, since solutions with the deepest smooth throats are on the cusp of becoming black holes.

The scaling solutions for multicenter Gibbons-Hawking metrics outlined in section~\ref{sec:MicroGeom} provide a strong test of the ideas of this paper, if one can understand enough about the dynamics in the regime where branes wrapping the small cycles on the base $\BB$ become light.  This is the regime where excitations of the long string become light and take over the effective dynamics -- classically the throat where it resides can grow infinitely deep and the string is naively tensionless as seen from the asymptotic region, in supergravity.  
As in the fivebrane case, one does not expect the long string to actually become tensionless, rather that its tension is small but finite as in the case of little string theory, related to the amount of fractionation of the fundamental string tension that it exhibits.
It would be helpful to know how the excitation gap arises once these degrees of freedom are included in the effective description.

We have seen that the new light degrees of freedom that are bound to the long effective string are visible in the regime where the `Coulomb branch' joins the `Higgs branch' of the underlying nonperturbative CFT, to borrow the terminology of quiver dynamics.  While the Higgs branch dynamics is strongly coupled and non-geometrical, and seems likely to carry the bulk of the entropy, we may be able to infer certain characteristics of black holes from the characteristics of the breakdown of the Coulomb branch description embodied by the microstate geometries.  

Such an approach was used successfully in%
~\cite{Horowitz:1997fr,%
Li:1998ci%
}
to find the scaling properties of black holes in matrix theory.
The starting point there was the Coulomb branch effective action for the zero modes of $N$ D-branes on a torus of size $L$
\be\label{oneloop}{
  \LL_{{\rm eff}}=\sum_{a=1}^N\frac{Nv_a^2}{R} + \sum_{a\ne b} \frac{N^2\lpl^9 |v_a-v_b|^4}{R^3L^d\,r_{ab}^{D-4}}+\dots\ .
}\ee
obtained by integrating out the strings stretching between branes.  Assuming the degrees of freedom lie in a region of size $r_0$ and saturate the uncertainty bound
\be\label{heisenberg}{
 \frac{r_0 v}{ R}\sim 1\ ,
}\ee
and applying the virial theorem, one arrives at a relation between the number of D-particles $N$ 
and the characteristic size $r_0$ of the bound state:
\be\label{Nsize}{
  N\sim (\lpl^{-9}L^d)r_0^{D-2}\ .
}\ee
The typical energy scale is then
\be\label{Emf}{
  E_{\rm lc}\sim (\lpl^{-9}L^dR)r_0^{D-4} =\frac{M^2R}{N}\ ,
}\ee
which is interpreted as the light-cone frame energy $P^-$ of a Schwarzschild black hole highly boosted to a momentum $P^+ = N/R$.
These considerations lead to a typical size of the bound state in terms
of the rest mass:
\be\label{msize}{
  M\sim (\lpl^{-9}L^d) r_0^{D-3}\ .
}\ee
Since $\lpl^{-9}L^d=1/G_D$, where $G_D$ is the D-dimensional Newton constant,
one finds the scaling relation between the
mass and horizon radius of a Schwarzschild black hole.
Using \pref{Nsize}, \pref{msize}, one also has
\be\label{Smf}{
  S\sim (\lpl^{-9}L^d)^{\frac1{D-3}} M^{\frac{D-2}{D-3}}
   \sim \lpl^{-9}L^d r_0^{D-2}\sim N\ .
}\ee
This result is already clear from \pref{Nsize} -- the number of D-particles
is the surface area of the bound state in Planck units.
In other words, the entropy is the number of constituent D-particles
up to coefficients of order unity.
This is quite reasonable
since they by assumption saturate the uncertainty bound,
and so $N$ is the number of phase space cells occupied by the system.

Taking into account the number of polarization states
for each D-particle, one estimates the entropy to be $S\sim N$.%
\footnote{One can remove the constraint that the entropy is tied to a particular choice of boost of the black hole by replacing the individual D0-branes in the above analysis with the motion of threshold bound states of $D0$ branes; see~\cite{Li:1998ci} for this and other generalizations.}
Similar considerations provide a picture of Hawking radiation as the emergence of D0-branes back onto the Coulomb branch~\cite{Banks:1997cm}.
Notice that this argument 
works uniformly in all dimensions $D$, and does not require
independent conjectures about the gauge theory thermodynamics.
The basic assumptions are simply (1) the Coulomb branch effective field theory~\pref{oneloop} is applicable (even if nearing breakdown);
and (2) the system is in a minimal uncertainty bound state.   

The (admittedly crude) picture just outlined approximates a Schwarzschild black hole in terms of the interactions of D-particle bound states.  In the simplest situation where the D-particles are D0-branes, the constituents are essentially a bundle of 11d gravitons travelling along null geodesics, and the interaction term in~\pref{oneloop} approximates the geodesic deviations of the bundle.  If one tries to localize that bundle too closely in the space transverse to the null trajectory, one finds that stretched string/membrane interactions among the gravitons are excited that disorder the light-cones, making the resulting trajectories rather non-commutative, chaotic, and quantum mechanically spread out (from the perspective of an outside observer).

This model for black holes in the D0-brane matrix model is not all that far removed from the picture of three-charge black holes advocated here.  The threshold bound state of $N$ D0-branes is a null wave which classically has a null singularity at its center.  Sending in a disturbance excites new light degrees of freedom near the singularity and sets up a cloud of such excitations extending out to the horizon radius.  The region of support of the D-particle wavefunctions seems quite similar to what one expects of the inter-horizon region of the three-charge system.  At a superficial level, the main difference is that the null singularity in the three-charge case lies at the boundary of an exterior region of low curvature which one expects to be well described by semi-classical gravity, whereas the region near the singularity of the extremal D-particle state has high curvature.

The matrix theory result shows that general principles can yield the scaling properties of the equation of state and the horizon size.  It is conceivable that enough could be pinned down about the effective theory of long strings near the black hole horizon that one could determine at least these same qualitative features of the dynamics, and compare with black hole thermodynamics.


\subsection{The information paradox, the experience of infall, \textbf{\textit{etc.}}}

Finally, let us turn to a discussion of causal structure, and the new ingredients provided by our scenario which are missing from typical discussions of the conflict between unitary evolution and causality in the context of black holes.  These typical discussions start with a sketch of the Penrose diagram of the classical geometry, and then proceed to a debate over how the information could possibly get out of the black hole, given that the geometry is smooth and semiclassical in the vicinity of the horizon where the Hawking process operates.  Current versions of the debate%
~\cite{Mathur:2009hf,Almheiri:2012rt} 
refine Hawking's original calculation by rephrasing the basic paradox in the language of quantum information theory.

The issue at its core is how to engineer the necessary correlations that carry quantum information over macroscopic spacelike distances, and preserve them from unwanted decoherence, while not proposing structures that do violence to cherished notions such as causality in contexts other than black hole dynamics.  An essential ingredient is likely to include the notion that causal structure in a theory of extended objects is quite tricky, and very likely not definable locally.  It has long been felt that the fact that the constituents of string theory are extended objects will play a vital role.  Any attempt to cleave the theory along the light cone structure of the low energy metric is doomed to failure, as for instance strings oscillate like mad even in their ground state; the zero-point fluctuations of the string oscillation guarantee that there are parts of the string on both sides of any imaginary dividing line.  Early investigations%
~\cite{%
Martinec:1993jq,%
Lowe:1994ns,%
Lowe:1995ac%
}
computed the commutator of string fields (admittedly an off-shell and not particularly gauge invariant quantity) and showed that light cones, defined as the boundary of the vanishing of the commutator, fuzz out due to string fluctuations.  However, it was never clear how this result would translate into a gauge invariant statement about how the notion of light cones determined by the effective gravity theory would be violated, or a specific mechanism for information retrieval from black holes, or how such a mechanism would not lead to unacceptable violations of causality in other contexts.  Indeed, it is quite remarkable that the tree level S-matrix of perturbative string theory satisfies all the usual analyticity properties required by causality, given how nonlocal strings seem to be.  One lesson that seems robust, however, is that the description of even a single extended object is highly entangled across the light cones of the effective geometry it inhabits.

The new ingredient provided by the emergence of long strings near the black hole phase transition, is that these strings lie at the correspondence principle crossover%
~\cite{Horowitz:1996nw}
where BTZ black holes turn into string states.  This issue has been studied in the context of perturbative string theory in $AdS_3$ backgrounds%
~\cite{Giveon:2005mi}, 
where one can vary the curvature of the $AdS$ geometry relative to the string scale by varying the superconformal field theory being coupled to the $AdS$ factor.  In terms of the level $k$ of the worldsheet $SL(2,\bR)$ CFT describing $AdS_3$, one has
\be
\label{NS5Radius}
\ell^2 = k \,\lstr^2 ~.
\ee
As one tunes through theories to go from curvature weaker than the string scale to curvature stronger than the string scale, the high energy spectrum crosses over from being dominated by BTZ black holes, to being dominated by perturbative strings.   The deep result of~\cite{Giveon:2005mi} is that beyond the crossover point, BTZ black hole states cease to be normalizable and therefore can't be excited because they are not part of the spectrum, {\it for any value of the mass}.  Precisely at the crossover, the BTZ spectrum matches the perturbative string spectrum, and one is at the correspondence point.  The difference with the original insight of~\cite{Horowitz:1996nw} is that in the latter work, the correspondence point occurs for one particular value of the mass that depends on the given value of the coupling; here it occurs for any value of the mass, but only for one particular coupling.  For smaller values of the coupling, there is no correspondence point, in fact there are no black holes at all.

In the analysis of~\cite{Giveon:2005mi}, this crossover occurs precisely where the string scale and the $AdS$ scale coincide, namely $k=1$.  But as that work emphasized, the crucial point is that the correspondence point is where the string spectrum matches the black hole spectrum.  This occurs almost by construction in $AdS_3/CFT_2$ duality, in which the long string density of states matches the BTZ spectrum exactly.  The entropy formula 
\be
S = 2\pi\sqrt{n_1n_5 (E+n_p)/2-J_L^2} + 2\pi\sqrt{n_1n_5(E-n_p)/2-J_R^2}
\ee
can either be interpreted as the density of states of BTZ black holes in a unitary theory of gravity in a weakly curved $AdS$ spacetime with $\ell = 4n_1n_5 G_3$, or as the density of states on a long string whose excitations have central charge $c=6$ and a tension reduced by a factor $n_1n_5$.
In this context, it is quite intriguing that the critical $k=1$ theory discussed in~\cite{Giveon:2005mi} has $c_{\rm eff}=6$.

It may thus happen that the degrees of freedom that hold the black hole entropy don't treat it as a black hole, because they don't see it -- they resolve the black hole singularity (in the sense of smoothness) by not resolving it (in the sense of measurement).  While short strings are experiencing horizons and singularities, the long string thinks that spacetime is smooth!  This proposition seems to be the logical extension of the results of~\cite{Giveon:2005mi}. 
It may not be such an outrageous proposition as it might seem at first -- we are used to different objects in string theory experiencing different metrics, see for instance~\cite{Gibbons:2000xe}.  
In the analysis of~\cite{Giveon:2005mi}, there are no black hole states in the spectrum beyond the correspondence point, just the string spectrum.   In the geometry that the long string responds to, there is no horizon and no singularity.  From this perspective, the long string resolves black hole singularities the way that perturbative strings resolve orbifold singularities -- by not feeling them.  In particular the long string will not respond to the ambient short string metric by falling into its singularity; instead, while short strings see a geometry which is locally $AdS_3$ with 
$\ell=4n_1n_5 G_3$ 
and a globally a BTZ black hole metric, long strings see $\ell_{\rm eff} = 4 G_{3,\rm eff}$ and behave entirely differently, in particular they see no black hole.

A very similar picture again arises for bound states of strings, fivebranes and momentum in a different limit.  The theory of ordinary `fundamental' strings in the throat of $n_5$ near-coincident NS5-branes is described by the worldsheet theories elaborated in%
~\cite{%
Aharony:1998ub,%
Giveon:1999zm,%
Giveon:1999px,%
Giveon:1999tq%
}. 
However, at the bottom of the throat lurks a nonperturbative `little string' whose tension is $n_5$ times smaller than that of the fundamental string, and it is this string that governs the thermodynamics%
~\cite{Maldacena:1996ya}
\be
S = 2\pi\sqrt{n_5 N_L-J_L^2} + 2\pi\sqrt{n_5N_R-J_R^2} ~.
\ee 
In the perturbative string theory description of NS5-branes, one has a throat with radius $\ell$ satisfying~\pref{NS5Radius} with $k=n_5$; little string theory has the tension reduced by a factor of $n_5$ and according to the entropy counting is at the correspondence point.
Despite their disparate names, `little strings' and `long strings' appear to be two sides of the same coin.

The work of~\cite{Giveon:2005mi} therefore provides a similar singularity resolution when fivebranes are the only `heavy' background charge.  Here, the short strings on the Coulomb branch see a capped throat with a linear dilaton described by $SL(2,\bR)/U(1)$ worldsheet conformal field theory at level $k=n_5$, which when sufficiently excited collapses to a linear dilaton black hole, described by the Lorentzian version of this same coset CFT.  The long string (or `little string') at the end of the throat has a tension $n_5$ times smaller, and so for it the throat geometry seems to have $k=1$; what it sees can equally well be described as a Liouville wall instead of a black hole.  The $SL(2,\bR)/U(1)$ sigma model has a strong/weak coupling duality to Liouville theory~\cite{Giveon:1999px} (see~\cite{Giveon:2001up} for a discussion and further references);
Liouville theory is the appropriate weakly coupled description for $k<1$, while the geometrical description is weakly coupled for $k>1$.  
For $k=1$, the black hole and Liouville wall are equally valid descriptions, but the Liouville version has no horizons or singularities, and we are free to use it.
Once again the long string `resolves' the singularity by not seeing it as such.
It is interesting that once again the worldsheet theory has $c_{\rm eff}=6$ as one would expect of the little string.

The large, floppy long/little string of exceedingly low tension will have a wavefunction that is coherent over macroscopic distances; and any attempt to decohere it through local measurements will fail, essentially because the large floppy string is a fault-tolerant structure of the sort seen in topological condensed matter systems~\cite{Kitaev:1997wr}~-- its information content is stored in a highly nonlocal fashion.  To determine the state of the long string would require the infalling observer to perform coherent measurements on scales of order the horizon size.  The gas of excitations of an extremely low tension string (having a truly tiny Hagedorn temperature) will be essentially impossible to detect for local observers, who will not be able to distinguish it locally from the vacuum.  A similar situation occurs in perturbative string theory when a D-particle enters the cloud of a highly excited fundamental string; its ballistic motion through the cloud is largely undisturbed over modest time scales.  And in the black hole problem, the appropriate time scale is set by the proper time of freely falling observers crossing the inter-horizon region.

The inter-horizon region is thus described by a coupled two-phase system -- a Hagedorn gas of the long string, weakly interacting with infalling ordinary strings. The experience of infall is expected to be smooth and uneventful until the observer hits the null singularity at the inner horizon.  The curvature singularity at the inner horizon is the signal that the coupling between short and long strings has grown large.  Tidal forces rip an infalling short string apart at the curvature singularity and fractionate it, at which point it has become absorbed into the long string sector.

The extremely light tension of the long string provides the sort of `nonviolent nonlocality'~\cite{Giddings:2012gc} that can provide an escape route for information to flow out of the black hole interior, again because the notion of locality is $n_1n_5$ times weaker for the long string than for short strings.  In this scenario, short strings pass freely through the ensemble geometry all the way to the inner horizon, where they are fractionated into the long string density of states and then gradually their information content is passed back into the short string spectrum in Hawking radiation outside the black hole as the long string decays back to extremality.  The long string responds to a different geometry, one that has no horizon or singularity, and thus has no difficulty communicating information in ways that short strings cannot.  And because outside of black hole regimes the long string is `confined', it will not do violence to cherished notions of locality and causality in other contexts.

\begin{figure}
\centerline{\includegraphics[width=5in]{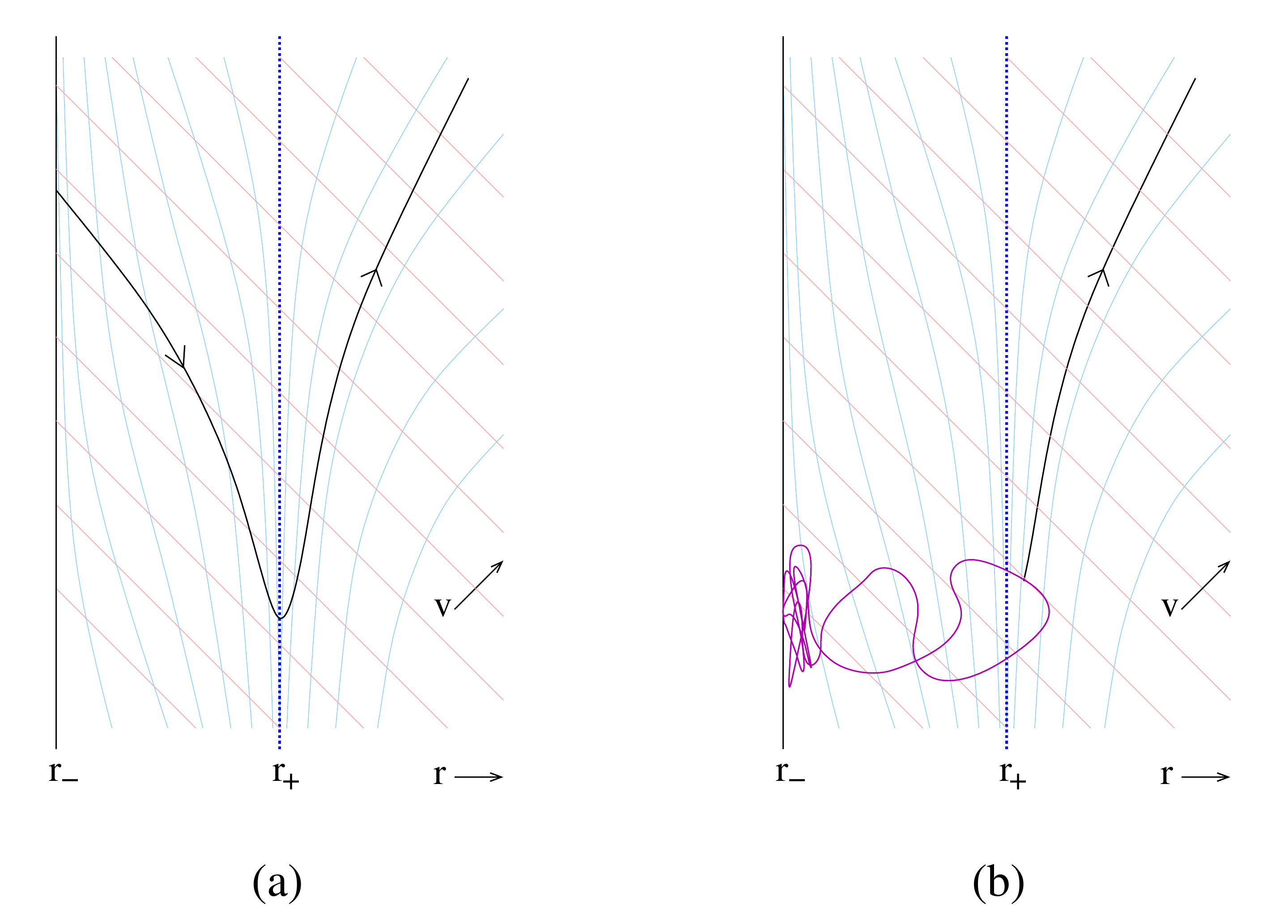}}
\setlength{\unitlength}{0.1\columnwidth}
\caption{\it 
Two descriptions of the Hawking process:  
(a) In the ensemble geometry, a particle traces a path from the singularity at the inner horizon backwards in time to the outer horizon, and then out to infinity; the part travelling backward in time is interpreted as the negative energy, antiparticle member of the Hawking pair produced at the horizon.  
(b) Radiation from the long string, whose degrees of freedom concentrate near the inner horizon of the not-too-nonextremal black hole but also extend out through the black hole interior to the vicinity of the outer horizon.  
}
\label{fig:HawkingRad}
\end{figure}

So what is missing in Hawking's calculation of black hole radiance?  In hindsight, it lacked a large, low-tension string in its Hagedorn regime, which interacts with the low-energy degrees of freedom, but which ignores the light-cone structure of the black hole geometry seen by those low-energy degrees of freedom.  When one traces over the long string degrees of freedom to obtain the ensemble geometry, one explicitly forgoes the ability to follow correlations between what fractionates into the long string sector when it hits the inner horizon and what escapes from the long string via Hawking radiation.  The description with the long string sector traced over seemingly has Hawking particles appearing randomly out of the vacuum, instead of being causally radiated by the long string.
In a path integral derivation of Hawking radiance such as%
~\cite{Hartle:1976tp}, one sums over all paths the particle could take from the future singularity to future null infinity $\II^+$, see figure~\ref{fig:HawkingRad}a.  The part running backwards in time from the future singularity to the future horizon is the path integral description of the antiparticle member of the Hawking pair created at the horizon.  Running the path to the singularity instead of having it connect to a vertex operator on the long string near the horizon, as in figure~\ref{fig:HawkingRad}b, misses the fact that information is conveyed from the singularity to the horizon by the very degrees of freedom one has traced over; instead, the antiparticle path cannot causally connect the radiated particle to anything inside the outer horizon, and so there is no way this procedure could have found anything but information loss.  The portion of the path backwards from the long string vertex to the horizon, and then along the antiparticle trajectory into the singularity, is an incorrect backward extrapolation by the asymptotic observer of where the particle came from -- an inappropriate substitute for the degrees of freedom that have been integrated out, which are inhabiting the black hole.%
\footnote{In particular, in the full theory, there is nothing particularly Planck scale going on other than at the singularity of the effective geometry.}

The picture of the fundamental origin of black hole radiance, as coherent radiation from a long string that carries the black hole entropy, dates back to the original calculations of~\cite{Callan:1996dv,Das:1996wn,Maldacena:1996ix}, which showed that at leading order in the deviation from extremality, the processes of absorption and emission from the long string using effective vertex operators could reproduce exactly the emission of low frequency Hawking quanta, including greybody factors.  One could have asked what happens to this picture of the Hawking process further from extremality, and how it connects to the effective geometry description.  The considerations above answer this question -- the long string is still present; it inhabits the interior of the black hole; it continues to carry the entropy; and it coherently emits the Hawking radiation.  It is perhaps not surprising that the same mechanism is in operation far from extremality; the major surprise is that in order for it to remain operative, the long string must react to the ambient short string geometry in such a different fashion than short strings do.  In order to get the information out of the interior of the black hole, the causal structure of the long string dynamics must be such that its degrees of freedom can float within the interior, and not collapse into a singularity like ordinary matter.  That different response to geometry appears to be responsible for both the resolution of the black hole singularity (as the place where short strings go to die and become fractionated into the long string), as well as the resolution of the puzzles and paradoxes of the flow of information in and out of black holes.

Our considerations make it natural to propose that the covariant entropy bound is giving us information about the support of the wavefunction of the long string degrees of freedom, in that the differential version of the bound tells us the distribution of those degrees of freedom in a given radial shell;%
\footnote{And that thus indeed the picture of the wavefunctions provided by the pure Higgs states of quiver quantum mechanics would be misleading.} the expression~\pref{LocalTemp} also gives the local temperature.
Furthermore, the Hawking process is a mean field calculation that describes the means by which short strings and long strings couple in the vicinity of the outer horizon, while unfortunately not keeping track of quantum correlations in the process.  
The D0-brane model sketched above points in the same direction -- that the support of the wavefunction of the accessible microstate degrees of freedom is the black hole interior, out to the outer horizon.  In hindsight, the covariant entropy bound applied to the black hole interior is trying to tell us that there are degrees of freedom supported in the inter-horizon region, that are not forced to head toward the singularity along with ordinary matter; these degrees of freedom are instead impervious to the demands of the light cone structure of the ensemble geometry, and instead float within the black hole and have their own internal clock related to the temperature.  Our proposal that the long string -- the object responsible for the entropy being counted by the covariant entropy bound -- lives at the correspondence point, provides a mechanism for how this could happen.  It is truly remarkable how general coordinate invariance of the effective theory keeps track of all the degrees of freedom present, no matter how hidden they are from those which are explicit in the effective theory.

As for the relation to exact dual CFT descriptions, it is of course hard to say given that we know little about the symmetric product orbifold CFT $(\bT^4)^N\!/S_N$ at strong coupling.  The coupling in this theory is a transposition twist operator in the symmetric group, whose role is to intertwine cycles.  At the orbifold point, wavefunctions are diagonal in a basis of words in the symmetric group; each word consists of a collection of cyclic permutations of length $n_i$ with $\sum_i n_i=N$.  The interaction, turned up to large values to get to the supergravity regime, can still be described in this basis but the basis will no longer diagonalize the Hamiltonian.  At the orbifold point, global $AdS$ is the ground state consisting of all cycles in the word being of the shortest possible length, while the black hole states are built on a single longest cycle whose length is of order~$N$.  In the interacting theory, it seems reasonable that spacetimes without a black hole will continue to be described by wavefunctions whose long cycle component is heavily suppressed, and the black hole transition is the Hagedorn transition where the long cycle sector opens up, and has significant support in the wavefunction, but all the time having a detailed balance between the various components of the wavefunction, which now include both short and long cycles.  One may imagine that, like an interacting string gas in the Hagedorn regime, in the black hole states there will be an `equilibrium' where the wavefunctions have both long cycles and short cycles in detailed balance, and that the short cycles describe supergravity in a weakly curved locally $AdS$ spacetime, while the long cycle describes the black hole states of the long string; and Hawking radiation is the transfer of information from the long cycle to the short cycles.  What is missing, because it is so difficult to extract bulk locality from this description, is a sense that the long cycle is by and large only inhabiting the inter-horizon region, and that the short cycles are also describing the inter-horizon region as well as the black hole exterior as they are seen by supergravity observables.

Finally, it would be intriguing to say the least if there were applications in cosmology of these sorts of two-phase systems of fractionated and non-fractionated objects interacting with one another.  Such a possibility has been explored by Verlinde~\cite{VerlindeTalk}, who suggests that one might think of dark energy and the low curvature of our universe as being due to the presence of a nearly tensionless fractionated brane state, whose tension is of order the horizon scale.  Related ideas on the origins of de~Sitter entropy have been explored by Silverstein in a series of works beginning with~\cite{Silverstein:2003jp}.

\begin{figure}
\centerline{\includegraphics[width=2.5in]{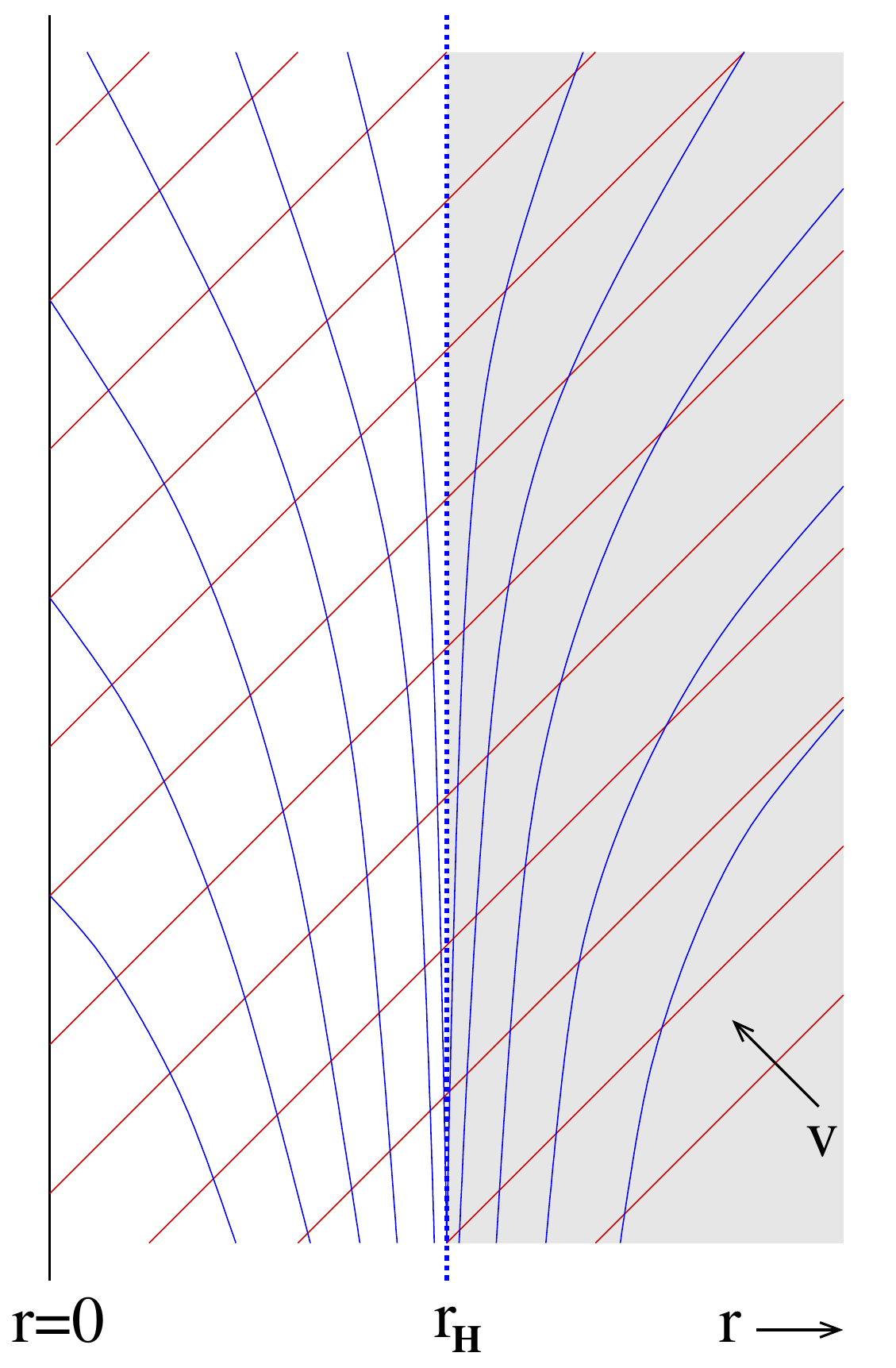}}
\setlength{\unitlength}{0.1\columnwidth}
\caption{\it 
Eddington-Finkelstein diagram for \ds spacetime.  The exterior $r>r_H$ of our Hubble volume is shaded.
}
\label{fig:deSitter}
\end{figure}

In the model advocated here, the entire picture of black holes is inverted in the cosmological context, see figure%
~\ref{fig:deSitter}.  
In this case we are living inside a bubble of metastable false vacuum, namely our Hubble volume, which is inhabited by short strings.  Instead of being outside the horizon looking in, we are inside the horizon looking out.  A generalization of the `long string state' made out of fractionated branes, \etc., inhabits the exterior of our Hubble volume; the two subsystems interact with one another in the vicinity of the de~Sitter horizon.  In this picture, de~Sitter symmetry would be a unitary symmetry transformation which acts to change the basis in the Hilbert space, moving some short string degrees of freedom into the fractionated brane sector and vice versa, thereby going to the frame appropriate to a different inertial observer.  One imagines that, as in the black hole context, the fractionated brane gas sees the geometry rather differently on distance scales less than the horizon size; it may also see a different light cone structure than that of the effective field theory, and be responsible for quantum coherence on super-horizon scales.

Rather than being an isolated system as in the black hole case, the fractionated brane gas now occupies all of space.  It should have a coherence scale, related to the horizon size and associated Hawking-\ds temperature.
Inflation is then a relaxation process, wherein the coherence length of the fractionated brane gas increases, its Hawking-\ds temperature drops, and the part of the fractionated sector accessible to the short string degrees of freedom grows with it; the Hubble volume increases in response to its co-evolution with the fractionated sector.  The analogue of Hawking radiation is the excitation of short string modes that are light compared to the Hubble scale.  The effective field theory interpretation of these fluctuations is that they are the evolution of the vacuum as modes are drawn up from below the Planck scale and then stretched to super-horizon scales; however, this would seem to be an incorrect extrapolation just as in the Hawking radiation case, with the effective field theory calculation being a stand-in for a more coherent and unitary process of radiation of short string modes by the fractionated brane gas.  These modes propagate out to larger radius, but in contrast to the black hole case this region is outside the region accessible to local observers.  During radiation or matter dominated eras, the fractionated brane gas relaxes much more rapidly, its coherence length grows, and mode fluctuations radiated during an earlier \ds era can re-enter the horizon.%
\footnote{Since the brane gas resolves spacelike singularities, there is no longer a reason to fear cosmological singularities, for instance as might occur if locally the brane gas exhibits a negative cosmological constant and the local universe of short strings collapses.  At the singularity, the short strings are merely reabsorbed into the fractionated brane gas.}

As the two-phase system of fractionated brane gas coupled to short strings relaxes, it can presumably get trapped in metastable minima; this is the landscape of string vacua seen by short strings.  Our currently accessible component of this vast system is quite near to `extremality'; this is the cosmological constant problem~-- to explain why our observed Hawking-\ds temperature is so low, given the presumably many other metastable minima the system can get trapped in where the short strings interact with many fewer available degrees of freedom of the fractionated brane gas.  Denef has been exploring analogous problems in the black hole context via ensembles of glassy brane bound states on the Coulomb branch, see for instance%
~\cite{Denef:2011ee,Anninos:2011vn,Anninos:2013mfa}.

Thus, perhaps the deepest mystery we currently face in cosmology is not the dark energy problem, but rather the \textbf{\textit{dark entropy}} problem -- why is essentially the entire entropy of the universe (\ie\ the area of our current cosmological horizon in Planck units) bound up in things we can't see?  From the perspective advocated here, we will not solve the riddle of dark energy without cracking the conundrum of dark entropy; and dark entropy -- in both black hole physics and in cosmology -- seems to have much to do with a sector of fractionated charges in string theory.



\vskip 2cm

\acknowledgments
The author is grateful to I. Bena and N. Warner for patient explanations of their work, to S. Mathur for extensive discussions, and to D. Marolf for an illuminating discussion of gravitational thermodynamics; 
thanks also to the organizers of the workshop {\it Quantum Information in Quantum Gravity} at UBC for hospitality during the course of this work.  E. Verlinde has expressed ideas relating his notion of `entropic gravity'%
~\cite{Verlinde:2010hp} to cosmology;
in a recent University of Chicago seminar~\cite{VerlindeTalk}, he proposed
a fractionated brane system filling our Hubble volume as a description of de~Sitter spacetime. 
This picture has a great deal of overlap with the discussion of cosmology in the final paragraphs above, and we thank him for communicating these ideas prior to publication.
This work was supported in part by DOE grant DE-FG02-13ER41958.

\vskip 2cm

\bibliographystyle{JHEP}
\bibliography{LongStrings}

\end{document}